\documentstyle[12pt]{article}  
\def\harr#1#2{\smash{\mathop{\hbox to .5in{\rightarrowfill}}\limits^{\scriptstyle#1}_{\scriptstyle#2}}}

\def\harrl#1#2{\smash{\mathop{\hbox to .5in{\leftarrowfill}}\limits^{\scriptstyle#1}_{\scriptstyle#2}}}

\def\varrow#1#2{\llap{$\scriptstyle #1$}\left\downarrow \vcenter to .5in {}\right.\rlap{$\scriptstyle #2$}}

\def\varrowu#1#2{\llap{$\scriptstyle #1$}\left\uparrow \vcenter to .5in {}\right.\rlap{$\scriptstyle #2$}}

\begin{document}
\title{\bf Differential forms, Hopf algebra and \\ General Relativity. I.}
\author{
J.F. Pleba\'nski$^{^1}$\thanks{e-mail:pleban@fis.cinvestav.mx}~, G.R. Moreno$^{^2}$\thanks{e-mail:gmoreno@math.cinvestav.mx}~,
F.J. Turrubiates$^{^1}$\thanks{e-mail:fturrub@fis.cinvestav.mx} \\[1ex]
{\it 1 Departamento de F\'{\i}sica}\\
{\it Centro de Investigaci\'on y de Estudios Avanzados del I.P.N.}\\
{\it Apdo. Post. 14-740, 07000, M\'exico, D.F., M\'exico.}\\[1ex]
{\it 2  Departamento de Matem\'aticas}\\
 {\it Centro de Investigaci\'on y de Estudios Avanzados del I.P.N.}\\
{\it Apdo. Post. 14-740, 07000, M\'exico, D.F., M\'exico.} } 
\date{}

\maketitle



\vspace{1cm}

\begin{abstract}

We review the language of differential forms and their applications to
Riemannian Geometry with an orientation to General Relativity. 
Working with the principal algebraic and differential operations on forms, 
we obtain the structure equations and their symmetries in terms of a new product 
(the co-multiplication).
It is showen how the Cartan - Grassmann algebra can be endowed with the structure of a
Hopf algebra.

\end{abstract}

\newpage

\begin{center}
\section{Introduction}
\end{center}
\setcounter{equation}{0}

This is the first part of two papers devoted to developing some new formalism in general relativity,
which is founded on the theory of differential forms and on the Hodge $*$-operation. 
\null

In order to work effectively in Newtonian theory we need the language of vectors; this
language condensed a set of three equations in one. But the most important of all
is that the vector formalism helps to solve problems in an easier way, and
furthermore this language reveals structure and offers insight.
The same occurs in the relativity theory, where the tensor language is needed, (again
the language helps to resume sets of equations and some structure in them). 

One of the principal advantages of classical vector analysis follow from
the fact that it enables one to express geometrical or physical relationships in a concise 
manner which does not depend on a coordinate system.
However, for many purposes of pure and applied mathematics or any other branch of the science the concept of vector is too 
limited in scope, (in the case of some basic geometrical or physical applications it is necessary to
introduce quantities which are more general than vectors), and in a significant extent, the
tensor calculus provides the appropriate generalization.
This also, has the advantage of a concise notation, and the formulation of its basic 
definitions is such as to allow an effortless work.

The calculus of differential forms (often called exterior calculus) represents a powerful tool 
of analysis whose use in mathematics and physics has become increasingly widespread. Like the 
tensor calculus its origins are to be found in differential geometry, largely as the result 
of the investigations of E.~Cartan towards the beginning of this century. 

The application of the differential forms in general relativity is now well known. Nevertheless, is seems to be interesting to show some new formalism in general relativity which is based on the application of the Hodge $*$-operation 
in theory of differential forms ~\cite{Ple},~\cite{Pleb2}. 
We suppose that this formalism enables one to understand in a better way the meaning of the Cartan structure equations in general relativity.

\begin{center}
\section{The Formalism of Scalar Components}
\end{center}
\setcounter{equation}{0}

This text uses two kinds of suffixes: $a,b,...= 1,...,n$ refer to the {\it rep\`eres} 
or forms, label them; and ${\alpha},{\beta},... = 1,...,n$ which denote the coordinates  
or tensorial indices. The summational convention applies with respect to the indices 
of all kinds. 
The construction considered is a construction over a differential manifold: 
the real $n$-dimensional differential manifold is denoted by $M_n$; 
a local map of real coordinates $x^{\alpha}$ is denoted
by $\{x^{\alpha}\}$.
The coordinates derivatives is denoted by comma: $T,_{\alpha}=\partial_{\alpha} T =
\partial{T}/\partial{x^\alpha}$.

The base of 1-forms on $M_n$ is denoted by $e^a$ : in a local 
map $\{x^{\alpha}\} : e^a = e{^a}_\mu dx^\mu \in \Lambda^1, e := det({e^a}_\mu)\neq 0$.
The vectors ${e^a}_\mu$ form the (local) co-base of all co-vectors. 
The space $\Lambda^1$ is a vector space with the elements being the 
functions of the point of $M_n$, which in $\{x^{\alpha}\}$  have the local 
representation: $\Lambda^1$ $\ni$ $\alpha = \alpha_{\mu} dx^\mu$, $\alpha_{\mu}(x)$ being the components 
of a co-vector with respect to $\{x^{\alpha}\}$. The vectors from $\Lambda^1$ called 1-forms are considered as 
defined over whole $M_n$; clearly the scalars  $\alpha_{\mu} dx^\mu$ do not depend on the 
choice of $\{x^{\alpha}\}$. 

Now, given in $\{x^{\alpha}\}$ co-base ${e^a}_{\mu}$ defines the contra-base ${e^\mu}_{a}$ 
by one of the two equivalent relations  
\begin{equation}
\label{1.1}
                   {e^a}_{\sigma} {e^\sigma}_{b}  =  {\delta^a}_{b},
              \,\,\,\,     {e^\alpha}_{s} {e^s}_{\beta} = {\delta^\alpha}_{\beta}.
\end{equation}
Due to (\ref{1.1}): $\Lambda^1$ $\ni$ $\alpha = \alpha_{\mu} dx^\mu$ = $\alpha_{a}e^a$; $\alpha_{a}
= \alpha_{\mu}{e_a}^{\mu}$.
This is the representation of an arbitrary vector from $\Lambda^1$ as spanned by 
the base of 1-forms. While the (local) co-base is induced by the (global) 
base of 1-forms, the contra-base is induced by the directional derivatives: $\partial_a = {e^\mu}_a \partial_\mu$.
We will use the notation $\partial_a T = T,_a$; of course $dT = dx^\alpha \partial_{\alpha}T =
e^a\partial_a{T}$.
With $e^a$ understood as defined on whole $M_n$, this relation provides the global definition of
the operators $\partial_a$.
It is clear that $e^a$ and $\partial_a$ are determined with the precision up to the 
transformations
\begin{equation} 
\label{1.2} 
      e^{a'}= {T^{a'}}_b e^b \,\, , \,\,    \partial_{a'}= {(T^{-1})^b}_{a'}\partial_a,
\end{equation}       
where $\|{T^a}_b\| \in GL(n,R)$, the parameters of $GL$ understood as regular functions on $M_n$.

Now, intending to define the Riemannian structure on $M_n$, we introduce the 
signature metric: it is a numeric, real symmetric non-singular $n{\times}n$ 
matrix: $g_{ab} = g_{(ab)}$, $det(g_{ab})\neq 0$, $g_{ab;c} = 0$ with the inverse $g^{ab}$. There exists 
such a $GL$ matrix that
\begin{equation} 
\label{1.3} 
      \|g_{cd}{T^c}_a{T^d}_b\| = \|diag(1...1,-1...-1)\| := \|\eta_{ab}\|,
\end{equation}    
with +1 appearing $n_{(+)}$, -1 appearing $n_{(-)}$ times; of course, $n_{(+)}+n_{(-)} = n$.
The pair of numbers $n_{(+)},\, n_{(-)}$ determines the dimension and the signature of a Riemannian 
space. Although one can work assuming from the very beginning $g_{ab} = \eta_{ab}$, we choose to work 
with the diagonalizable $g_{ab}$ but not necessarily of the diagonal form.
Now, the Riemannian space $V_n$ is the manifold $M_n$ equipped with the Riemannian metric.
The Riemannian metric of signature  $n_{(+)},\, n_{(-)}$ is the symmetric non-singular tensor
of valence 2, given in $\{x^{\alpha}\}$ by
\begin{equation}  
\label{1.4}
      g_{\mu\nu} := g_{ab}{e^a}_{\mu}{e^b}_\nu .
\end{equation}    
It induces the interval $ds^2 := g_{\mu\nu}dx^{\mu}dx^{\nu} = g_{ab}e^ae^b$. Thus, we accept the 
point of view that $g_{ab}$ and some base of 1-forms, $e^a$, induce the Riemannian metric.
It is clear that with $ds^2$ fixed, the forms  $e^a$ remain arbitrary only with respect 
to a subgroup of $GL$
\begin{equation}
\label{1.5}  
      e^{a'} = {L^{a'}}_b e^b ,\,\,\,\, g_{ab}{L^a}_c{L^b}_d = g_{cd}.
\end{equation}    
This subgroup coincides with $O(n_{(+)},n_{(-)})$, the transformations (\ref{1.5}) cannot affect metrical 
concepts; we will call them ``the metrical gauge" or ``the $O_n$ gauge".

Although we assume the tensor calculus as known, we should like now to 
present a brief summary of basic tensorial formula: our purpose is to fix the 
notational conventions which can slightly differ in the texts by different authors 
~\cite{Wald} -~\cite{Eguchi}.

The Christoffel symbols $\{^{\,\alpha}_{\beta \gamma}\}:=
\frac{1}{2}g^{\alpha\rho}(g_{\beta\rho,\gamma}+g_{\gamma\rho,\beta}-g_{\beta\gamma,\rho})$  
define the covariant differentiation $``;_\alpha"$ - in its sense $g_{\alpha \beta ; \gamma}=0$-.

Remember then, that directly from the transformation laws the sum of two connections is not a
connection or a tensor. However, the difference of two connections is a tensor of valence (1,2),
because the inhomogeneous term cancels out in the transformation. For this reason the
anti-symmetric part of a $\{^{\,\alpha}_{\beta \gamma}\}$, namely,
\begin{equation}
\label{1.15b}
{T^\alpha}_{\beta \gamma}=\{^{\,\alpha}_{\beta \gamma}\}-\{^{\,\alpha}_{\gamma \beta}\},
\end {equation} 
is a tensor called {\it the torsion tensor}. If the torsion tensor vanishes, then the connection 
is symmetric , {\it i.e.} 
\begin{equation}
\label{1.15c}
\{^{\,\alpha}_{\beta \gamma}\} = \{^{\,\alpha}_{\gamma \beta}\}.
\end{equation}
From now on, we shall restrict ourselves to symmetric connections.

The curvature tensor:
\begin{equation} 
\label{1.6} 
   R^\alpha_{\beta\gamma\delta}=-\{^{\,\,\alpha}_{\beta \gamma}\}_{,\delta} + \{^{\,\,\alpha}_{\beta
   \delta}\}_{,\gamma}+ \{^{\,\,\alpha}_{\sigma \gamma}\}\{^{\,\,\sigma}_{\beta \delta}\} -
   \{^{\,\,\alpha}_{\sigma \delta}\}\{^{\,\,\sigma}_{\beta \gamma}\},
\end{equation}    
has the basic independent symmetries $R_{\alpha\beta\gamma\delta}= R_{[\alpha\beta]\gamma\delta} =
R_{\alpha\beta[\gamma\delta]}, R_{\alpha[\beta\gamma\delta]}= 0,$ (where for all the
expressions from now on, the indices between brackets indicate the alternating sum over all
permutations of the indices),
so that it has $(^n_2)^2-n(^n_3) = \frac{n^2(n^2-1)}{12}$ of independent components. The basic symmetries
imply the secondary symmetry $R_{\alpha\beta\gamma\delta}= R_{\gamma\delta\alpha\beta}$.
The curvature tensor defines the secondary Ricci tensor $R_{\alpha\beta} = R^\rho_{\alpha\rho\beta}$ and the scalar 
curvature $R = R^{\rho}_{\rho}$; the Einstein tensor is defined by $G^\alpha_\beta = R^\alpha_\beta -\frac{1}{2}\delta^\alpha_\beta R$; the 
Ricci tensor with the extracted trace will be denoted $\overline{R^\alpha_\beta}=R^\alpha_\beta - \frac{1}{n}\delta^\alpha_\beta R$.

The conformal curvature or the Weyl's tensor $C^\alpha_{\beta\gamma\delta}$ is defined by
\begin{equation} 
\label{1.7}  
     {R^{\alpha \beta}}_{\gamma \delta}= {C^{\alpha \beta}}_{\gamma \delta} - \frac{1}{n-2} \delta^{\alpha \beta
     \rho}_{\gamma \delta \sigma}\overline{R^\sigma_\rho} + \frac{1}{n(n-1)} \delta^{\alpha \beta}_{\gamma
     \delta}R.
\end{equation}    
The Weyl's tensor has all symmetries of the curvature tensor, besides all its traces 
vanish; it can be no trivial for $n{\geq}4$. The covariant derivatives of 
the curvature tensor fulfill the Bianchi identities 
\begin{equation}  
\label{1.8a}
       R^\alpha_{\beta[\gamma\delta;\epsilon]} = 0 \rightarrow  G^\alpha_{\beta;\alpha}=0.
\end{equation}    
The relations (\ref{1.8a}) are called the special Bianchi identities. Notice that (\ref{1.7})
combined with (\ref{1.8a}) gives:
\begin{equation} 
\label{1.9} 
     {C^{\alpha \beta}}_{[\gamma \delta ; \epsilon]} - \frac{1}{3} \frac{1}{n-2} \delta^{\alpha \beta \nu
     \rho}_{\gamma \delta \epsilon \sigma} \left[R^\sigma_\rho - \frac{1}{2} \frac{1}{n(n-1)}
     \delta^{\sigma}_{\rho}R \right]_{;\nu}=
     0.
\end{equation}    
[For $n > 3$, (\ref{1.9}) $\rightarrow$ (\ref{1.8a})]. The generalized Kronecker's symbols which
appear in (\ref{1.7}) or (\ref{1.9}) can be defined by
\begin{equation}  
\label{1.10a} 
  \delta^{\alpha_1..\alpha_p}_{\beta_1..\beta_p} =
  \frac{1}{(n-p)!}\epsilon^{\alpha_1...\alpha_p\sigma_1...\sigma_{p}}\epsilon_{\beta_1...\beta_p\sigma_1...\sigma_p},
\end{equation}    
where $\epsilon's$ are Levi-Civita's densities, or the familiar determinant expression. The parallel
identity
\begin{equation} 
\label{1.10b} 
      \delta^{a_1..a_p}_{b_1..b_p} =
      \frac{1}{(n-p)!}\epsilon^{a_1...a_{p}s_1...s_{p}}\epsilon_{b_1...b_{p}s_1...s_p},
\end{equation}    
will be useful in our further considerations.

Now, if $T^{\alpha...}_{\beta...}$ are components of a tensor density of the weight $\omega$ given 
in $\{x^{\alpha}\}$, then we can define its scalar components by the formula:
\begin{equation} 
\label{1.11} 
      T^{a...}_{b...}:= e^{\omega}{e^a}_{\alpha}{e_b}^{\beta}...T^{\alpha...}_{\beta...}~.
\end{equation}    
The density of the weight $\omega = 1, \epsilon_{\alpha_1...\alpha_n}$ (this statement fixes our conventions in defining
weights) has the scalar components $\epsilon_{a_1...a_n}$. The scalar components given as functions of
the point of $M_n$ plus the known base of 1-forms give the global definition of a field of the
tensorial density over $M_n$, independent of the choice of the local maps.

Now, one easily finds that
\begin{equation}
\label{1.12}  
      T^{a...}_{b...;c}:= e^{\omega}{e^a}_{\alpha}{e_b}^{\beta}...T^{\alpha...}_{\beta...;\gamma}{e_c}^\gamma = 
      T^{a...}_{b...,c}+ {\Gamma^a}_{sc}T^{s...}_{b...}-{\Gamma^s}_{bc}T^{a...}_{s...}+...
\end{equation}    
where
\begin{equation}
\label{1.13}  
        {\Gamma^a}_{bc} := -{e^a}_{\mu;\nu}{e^\mu}_{b}{e^\nu}_c,
\end{equation}    
are the Ricci rotation coefficients. These can be understood as defined by Ricci forms
\begin{equation} 
\label{1.14} 
      {\Gamma^a}_b := {\Gamma^a}_{bc}e^c = 
      -{e^a}_{\mu;\nu}{e_b}^{\mu}dx^{\nu} \in \Lambda^1,
\end{equation}    
interpreted as the objects given on whole $M_n$. Notice that in the result of the $O_n$ gauge
these forms transform according to
\begin{equation}
\label{1.15}  
      {\Gamma^{a'}}_{b'} = {L^{a'}}_{a}{(L^{-1})^b}_{b'}{\Gamma^a}_b -
      {(L^{-1})^s}_{b'}{dL^{a'}}_{s}.
\end{equation}   
Denoting $\Gamma_{abc} = g_{as}{\Gamma^s}_{bc}$, one infers from $g_{ab;c} = 0$ that
\begin{equation} 
\label{1.16} 
       \Gamma_{abc} = \Gamma_{[ab]c},
\end{equation}    
and therefore
\begin{equation}  
\label{1.17}
      \Gamma_{abc} = \Gamma_{a[bc]}+\Gamma_{b[ac]} - \Gamma_{c[ab]}.
\end{equation}    
Moreover, ${\Gamma^a}_{bc}$ can be algebraically constructed from
\begin{equation} 
\label{1.18} 
      \Gamma^a_{[bc]} = -e^a_{[\mu,\nu]}{e^\mu}_b{e^\nu}_c,
\end{equation}    
the objects constructed from the ``rotations" $e^{a}_{[\mu,\nu]}$, the objects which do not involve the use
of the covariant derivatives.
Notice that these objects determine the commutator of the directional derivatives
\begin{equation} 
\label{1.19} 
      T^{...}_{...,[cd]} = T^{...}_{...,s}\Gamma^s_{[cd]}.
\end{equation}    
The objects $\Gamma^a_{[bc]}$ can be also interpreted in terms of the concept of the Lie bracket. In the
tensor calculus one introduces the Lie bracket of a pair of vectors given in $\{x^{\alpha}\}$ by the
components $\alpha^\mu,\beta^\mu$ as the vector ~\cite{Dub},~\cite{Gug},~\cite{Curtis}
\begin{equation} 
\label{1.20} 
      [\alpha,\beta]^\mu =
      \alpha^\sigma{\beta^\mu}_{,\sigma}-\beta^\sigma{\alpha^\mu}_{,\sigma}
      =\alpha^\sigma{\beta^\mu}_{;\sigma}-\beta^\sigma{\alpha^\mu}_{;\sigma}.
\end{equation}    
Consider now the 1 - forms
\begin{equation}  
\label{1.21}
      [\alpha,\beta] := [\alpha,\beta]_\mu dx^\mu = 
      (\alpha_\sigma{\beta_\mu}^{;\sigma}-\beta_\sigma{\alpha_\mu}^{;\sigma})dx^\mu \,\,\, \in 
      \Lambda^1.
\end{equation}    
Then the operation $[\alpha,\beta]$ can be considered as the map of an ordered pair $\alpha = \alpha_\mu dx^\mu, \beta =
\beta_\mu dx^\mu \in \Lambda^1$ into the same space $\Lambda^1$ $i.e.$ [$\cdot,\cdot$] :$ \Lambda^1 \times \Lambda^1 \rightarrow \Lambda^1$. The map [$\cdot,\cdot$] is of course bi-linear and has the basic properties of the Lie - composition
\begin{equation}
\label{1.22a}  
      [\alpha,\beta] + [\beta,\alpha] = 0,
\end{equation}    
\begin{equation} 
\label{1.22b} 
      [\alpha,[\beta,\gamma]]+[\beta,[\gamma,\alpha]]+[\gamma,[\alpha,\beta]] = 0.
\end{equation}    
Now, one easily sees that with $\alpha = \alpha_ae^a, \, \beta = \beta_ae^a$ we have
\begin{equation}
\label{1.23}  
      [\alpha,\beta] =
      (\alpha^s\beta_{a,s}-\beta^s\alpha_{a,s}+2\Gamma_a^{[rs]}\alpha_{r}\beta_s)e^a,
\end{equation}    
where the indices are manipulated by the signature metric.
In particular:
\begin{equation}  
\label{1.24}
      [e^a,e^b] = 2e^s\Gamma_s^{[ab]}. 
\end{equation}    

Of course, this formula can be interpreted as a global statement.
The formalism which works with scalar components and Ricci rotation coefficients is of
course alternative to the standard tensorial calculus which works with the local components
and the Christoffel symbols. Working with the scalar components one encounters a slightly
more complicated situation: the directional derivatives do not commute while the coordinates
derivatives commute. On the other hand, the Ricci rotation coefficients equipped with the
symmetry $\Gamma_{abc} = \Gamma_{[ab]c}$ are more convenient  - even in the local considerations - than the
Christoffel symbols with the symmetry $\{^{ \alpha}_{\beta\gamma}\}=\{^{
\alpha}_{\gamma\beta}\}$. While the second
has in general  $n(^{n+1}_{\,\,\,2})$ of independent components, the first posses in general only $n(^n_2)$ of 
independent components.
Furthermore, a formalism independent explicity on the choice of the local coordinates
(equipped in the remaining freedom of choice of the metrical gauge) is very useful when one is
interested in the global aspects of the metrical geometry.

For these reasons, in these days, it is very common to execute even the practical computation in
general relativity working with scalar components and the Ricci rotation coefficients. The
base of 1-forms one selects -within the freedom of the $O_n$ gauge - so this base becames
correlated with the vector fields with the geometric and physical interpretation present
in the theory. The practical computations show that this technique is convenient indeed,
that using it one can arrive to the final result more quickly than when operating with the
standard tensorial techniques ~\cite{Wald},~\cite{Kram},~\cite{Chan},~\cite{Misner},~\cite{Eguchi}.

\begin{center}
\section{Differential forms}
\end{center}
\setcounter{equation}{0}
\subsection{Forms}
In the first section we encountered  ready the space of $1$ - forms $\Lambda^1$, as introduced on the intuitive
level. Now, there exist various abstract and elegant definitions of the general $p$ - forms and the
corresponding {\it Cartan - Grassmann algebra} at a point $P \in M_n$. For our purposes,
however, it is sufficient to outline the basic ideas of the corresponding constructions from the
heuristic point of view only. The reader interested in the rigorous mathematical formulation could find
it in the literature ~\cite{GS},~\cite{Flan},~\cite{Schutz},~\cite{Schreiber},~\cite{Pleb1}.

The Cartan algebra $\Lambda$ is the set of the formal sums with the local representation in $\{x^{\alpha}\}$
\begin{equation} 
\label{2.1} 
      \Lambda \ni \alpha = \sum^n_{p=0} \alpha_{\mu_1...\mu_p}(x) dx^{\mu_1}\wedge ... \wedge dx^{\mu_p},
\end{equation}    
where  $\alpha_{\mu_1...\mu_p}= \alpha_{[\mu_1...\mu_p]}$ are $\{x^{\alpha}\}$ components of a sequence of the totally skew tensors of all
possible valences. The sequence of the abstract elements is denoted
\begin{equation}
\label{2.2}  
     1, ~ dx^{\alpha_1}, ~ dx^{\alpha_1} \wedge dx^{\alpha_2},~...~, ~
     dx^{\alpha_1} \wedge .... \wedge dx^{\alpha_n},
\end{equation}    
and represents the base of $\Lambda$ interpreted as a linear vector space. Thus, the operations $\alpha$ +
$\beta$, $c\alpha$, $c \in {\bf R}$ have the natural definitions as the operations on the coefficients of the
base. Consistently with $\alpha_{\mu_1..\mu_p} = \alpha_{[\mu_1..\mu_p]}$  it is natural to introduce an identification among the elements of
the base (\ref{2.2}), namely, to assume that: $dx^{\alpha_1} \wedge ... \wedge dx^{\alpha_p} = dx^{[\alpha_1} \wedge
... \wedge
dx^{\alpha_p]}$.
This means in particular that $dx^{\alpha_1} \wedge ... \wedge dx^{\alpha_2} = - dx^{\alpha_2} \wedge
... \wedge
dx^{\alpha_1}$ for an odd number of permutations. The last remark suggests how to make $\Lambda$ an algebra
by defining in it a map $\Lambda$ $\times$ $\Lambda$ $\rightarrow$ $\Lambda$.
This map we will denote $\wedge$ and we will call it the external multiplication.

First of all, the $\wedge$ has to be bi-linear, {\it i.e.}, $(\alpha +
\beta)\wedge\gamma = \alpha\wedge\gamma + \beta\wedge\gamma$  and
$\alpha\wedge(\beta + \gamma)=\alpha\wedge\beta + \alpha\wedge\gamma$,  for every $\alpha , \beta , \gamma \in
\Lambda$. Therefore, to
determine the map $\wedge$, it is enough to define how $\wedge$ works as applied with respect to the
elements of the base (\ref{2.2}). This we fix by assuming the three consistent properties:

\begin{itemize}
\item[1)] $c \wedge dx^\alpha = dx^\alpha \wedge c = cdx^\alpha \in \Lambda^1, c\in {\bf R}$; in particular $1\wedge dx^\alpha = dx^\alpha \wedge 1 = dx^\alpha$; this will 
assure that $\Lambda$ is an algebra with unit.

\item[2)] $(dx^{\alpha_1}\wedge...\wedge dx^{\alpha_p})\wedge(dx^{\beta_1}\wedge...\wedge
dx^{\beta_p}) = dx^{\alpha_1}\wedge...\wedge dx^{\alpha_p} \wedge dx^{\beta_1}\wedge...\wedge
dx^{\beta_p}$. This can be interpreted as the associativity of the $\wedge$ composition of
differentials.

\item[3)] $(dx^{\alpha_1})\wedge (dx^{\alpha_2}) = -(dx^{\alpha_2})\wedge (dx^{\alpha_1})$; this skewness property together 
with 2) implies $dx^{\alpha_1}\wedge...\wedge dx^{\alpha_p} = dx^{[\alpha_1}\wedge...\wedge dx^{\alpha_p]}$ what must hold for consistence anyway.
\end{itemize}
Notice that due 3) $p + q > n  \,\,\, \rightarrow (dx^{\alpha_1}\wedge...\wedge dx^{\alpha_p})\wedge(dx^{\beta_1}\wedge...\wedge
dx^{\beta_q}) = 0$  [zero in the sense of $\Lambda$ interpreted as a
vectorial space].

\null

The algebra $\Lambda$ equipped in the operations +, $\wedge$, from the local point of view represents
the set of sequences of the skew tensors of all possible valences, equipped in the corresponding
algebraic operations on these sequences; $\alpha + \beta$ corresponds to the addition of the elements of
the same order, $\alpha_{\mu_1...\mu_p}+ \beta_{\mu_1...\mu_p}$ while $\alpha \wedge \beta$ is characterized 
by the sequence : \newline $\sum^p_{q=0}\alpha_{[\mu_1...\mu_{p-q}}\beta_{\mu_{p-q+1}...\mu_p]}$.

Now, the important point consists in the fact that $\alpha \in \Lambda$ can be interpreted as a sequence of the skew
tensors defined invariantly over whole $M_n$, independently of any local maps.

Indeed, using the base of 1- forms introduced in the previous section and the actual assumptions concerning the
external multiplication $\wedge$, one easily sees that the typical element of $\Lambda$,
(\ref{2.1}) can be represented as
\begin{equation}
\label{2.3}  
      \Lambda \ni \alpha = \sum^n_{p=0} \alpha_{a_1...a_p}e^{a_1}\wedge ... \wedge e^{a_p},
\end{equation}
where $\alpha_{a_1...a_p}= \alpha_{[a_1...a_p]}$ are the scalar components of the tensors from
(\ref{2.1}) and $e^{a_1}\wedge...\wedge e^{a_p}$ is the external product of the 1-forms
which constitute the base; it is obvious that so interpreted $\alpha$ dependent on the point of $M_n$ is defined over
whole $M_n$.
Of course, $\alpha$ from (\ref{2.1}) [or (\ref{2.3})] can be regarded as $\alpha = \sum^n_{p=0}\alpha_p$ where $\alpha_p$ is that part of $\alpha$ which is
spanned on the external product of $p$ differentials [$p$ basic 1 - forms]. These homogeneous elements as such form a
linear vector space denoted $\Lambda^p$ and are called $p$-forms.
When dealing with $\alpha_p$ belonging to a definite $\Lambda^p$ we will omit the suffix $p$, $\alpha = \alpha_p$
consistently with (\ref{2.1}).
If $\alpha \in \Lambda^p$ we will say that its degree is $p$, $deg(\alpha) = p$, while its co-degree $p' = n - p$,
$codeg(\alpha)= p'$.
Of course $deg(\alpha) + codeg(\alpha) = n$. A general $\alpha \in \Lambda$ has the degree (co-degree) indeterminated. It is
clear that $\Lambda = \bigoplus^n_{p=0} \Lambda^p$.

A general $p$-form has the standard representation:
\begin{equation}
\label{2.4}  
       \Lambda^p \, \ni \,\, \alpha = \alpha_{\mu_1...\mu_p}dx^{\mu_1} \wedge ... \wedge dx^{\mu_p} =
       \alpha_{a_1...a_p}e^{a_1}\wedge ... \wedge e^{a_p}.
\end{equation}
Of course, $n \geq p \geq 0, \alpha \in \Lambda^0$ are interpreted as functions on $M_n$. Sometimes it is convenient to work with
$p$-forms in a slightly different normalization ~\cite{GS},~\cite{Schutz}: we will denote
\begin{equation}  
\label{2.5}
      \Lambda^p \, \ni \,\, \buildrel{\sim}\over{\alpha} \,\, := \,\, \frac{1}{p!} \alpha = \frac{1}{p!} 
       \alpha_{a_1...a_p}e^{a_1}\wedge ... \wedge e^{a_p}.
\end{equation}
The external multiplication can be regarded as the bi-linear map:
\begin{equation} 
\label{2.6} 
       \wedge \,: \, \Lambda^p \times \Lambda^q \,\,\,\, \rightarrow \,\,\,\, \Lambda^{p+q},
\end{equation}
equipped in the properties
\begin{equation}
\label{2.7}  
      1 \wedge \alpha = \alpha \wedge 1 = \alpha, \,\,\, {\rm{for ~ every}} \, p,\,\, deg(\alpha) = p,
\end{equation}
\begin{equation}
\label{2.8}  
  deg(\alpha)=p \, , \, deg(\beta)=q \,\,\,\, \rightarrow \,\,\,\, \alpha \wedge \beta = (-1)^{pq}\beta \wedge
  \alpha,
\end{equation}
(also : $deg\alpha + deg\beta > n \,\, \rightarrow \,\, \alpha \wedge \beta = 0$) and
\begin{equation}
\label{2.9}
     (\alpha \wedge \beta) \wedge \gamma  \,\,= \,\,  \alpha \wedge (\beta \wedge \gamma), 
\end{equation}
for every $\alpha,\beta,\gamma$ of definite degrees. That means, that the next diagram
\begin{equation}
\label{2.9,1}
\begin{array}{ c c c }
   \Lambda \times \Lambda \times \Lambda &\harr{id \times \wedge}{}&\Lambda \times \Lambda\\
   \varrow{\wedge \times id}{}& &\varrow{\wedge}{}\\ 
  \Lambda \times \Lambda & \harr{\wedge}{} & \Lambda\\
  \end{array} 
\end{equation}
is commutative.

This map extended on the direct sum of all
$\Lambda^p$$'s$ becomes the external multiplication of $\Lambda$.

\subsection{Star operation}

Now, we will define the star operation $*$ (called sometimes the Hodge's operation ~\cite{Eguchi},~\cite{GS}, \cite{Flan}).
Its existence is possible only in the theory of oriented $M_n$ equipped with
the Riemannian metric. The
star operation is a map of $\Lambda$ onto $\Lambda$ which results from the 
``partial maps", $*\alpha = \sum^n_{p=0} \,\,*\alpha_p$ which act according to the scheme: 
\begin{equation}
\label{2.10}
      *\,\, :\,\, \Lambda^p \,\,\, \rightarrow \,\,\, \Lambda^{p'}  \,\,\,\,\, (p'=n-p),
\end{equation}
and is defined for $\alpha\in\Lambda^p$ with the representation (\ref{2.4}) by the formula
\begin{eqnarray}
\label{2.11}
        *\alpha & = & \frac{\mid det(g_{\alpha\beta}) \mid^{\frac{1}{2}}}{(n-p)!} {\epsilon^{\lambda_1...\lambda_p}}_
        {\mu_1...\mu_{n-p}}\alpha_{\lambda_1...\lambda_p}dx^{\mu_1}\wedge ... \wedge dx^{\mu_{n-p}},
        \nonumber \\ & = & 
         \frac{\mid det(g_{ab}) \mid^{\frac{1}{2}}}{(n-p)!} {\epsilon^{a_1...a_p}}_
        {b_1...b_{n-p}}\alpha_{a_1...a_p}e^{b_1}\wedge ... \wedge e^{b_{n-p}}. 
\end{eqnarray}
In the first line the indices of the Levi-Civita's symbol are manipulated by the 
Riemannian metric; in the second by the signature metric; both lines are equal provided:
\begin{equation}
\label{2.12}
     \mid det(g_{\alpha\beta}) \mid^{\frac{1}{2}} = +\mid det(g_{ab})\mid^{\frac{1}{2}} e,
\end{equation}
(Remember that $e := det({e^a}_\mu)$).

\null
 
This will be our standing convention concerning the choice of the branch for the
root of the determinant of the Riemannian metric.

The integration of a differential form in a manifold $M$ is valid when $M$ is orientable; we define an
orientation of a manifold like follows: Let $M$ be a connected $n$-dimensional differentiable manifold. At a point
$x \in M$, the tanget space $T_xM$ is spanned by the basis $\{e_\mu \}=\{\partial / \partial x_\mu \}$, where
$x^\mu$ are the local coordinates on the chart $U_i$ to which $x$ belongs. Let $U_j$ be another chart such that
$U_i \cap U_j \neq \emptyset$ with the local coordinates $y^\alpha$. If $x \in U_i \cap U_j$, $T_xM$ is spanned
by either $\{e_\mu \}$ or $\{ e'_\alpha \}= \{ \partial / \partial y^\alpha \}$. The basis changes as:
\begin{equation}
 e'_\alpha = (\partial x^\mu / \partial y^\alpha)e_\mu.
\end{equation}
If $J = det(\partial x^\mu / \partial y^\alpha) > 0$ on $U_i \cap U_j$, $\{e_\mu \}$ and $\{e'_\alpha\}$ are said
to define the $same \,\, orientation$ on $U_i \cap U_j$ and if $J<0$, the $opposite \,\, orientation$. 

Thus for a connected manifold $M$ covered by $\{U_i \}$, it is $orientable$ if, for any overlapping charts $U_i$
and $U_j$, there exist local coordinates $\{x^\mu \}$ for $U_i$ and $\{y^\alpha \}$ for $U_j$ such that
$J=det(\partial x^\mu / \partial y^\alpha) > 0$ ~\cite{Flan},~\cite{Nak},~\cite{Schwarz}. 

\null

The second line of (\ref{2.11}) emphasises the invariant nature of $*$ operation, that is, its 
independence of local maps.
The definition (\ref{2.11}) and (\ref{1.10b}) imply that 
$deg \alpha = p,~deg\beta = p+q, ~ n\geq p+q \geq 0$, then 
\begin{equation}
\label{2.13}
   *(\alpha\wedge*\beta) = (-1)^{qq'+n_{(-)}} \frac{(p+q)!}{q!}
   \beta_{a_1...a_qb_1...b_p}\alpha^{b_1...b_p}e^{a_1}\wedge...\wedge e^{a_q},
\end{equation}
[$(-1)^{n_{(-)}}$ corresponds to the factor $sign\,det(g_{ab})$]. This useful formula 
specialized for $\alpha\rightarrow 1,~p\rightarrow 0,~\beta\rightarrow\alpha,~q\rightarrow p$
gives
\begin{equation}
\label{2.14}
  deg(\alpha) = p \,\,\,\, \rightarrow \,\,\,\, **\alpha = (-1)^{pp'+n_{(-)}}\alpha.
\end{equation}
It follows that the operation $\star$ defined by:
\begin{equation}
\label{2.15a}
   deg(\alpha) = p {\,\,\,\,} \rightarrow {\,\,\,\,} \star\alpha = exp(\frac{i}{2}\pi[pp'+n_{(-)}])\cdot*\alpha
\end{equation}
has the property:
\begin{equation}
\label{2.15b}
   deg(\alpha) = p \,\,\,\, \rightarrow \,\,\,\, \star\star\alpha = \alpha.
\end{equation}
Therefore, $\star$ can be interpreted as an involutional automorphism of 
$\Lambda$. When one works with forms having the coefficients in ${\bf C}$ and the imaginary 
factors do not matter, it is advantageous to work with $\star$ instead of $*$. In 
the case of real forms, however, avoiding imaginary factors, it is customary 
to work rather with $*$ than with $\star$, so we will do in this text. One can add 
that due to $(-1)^{pp'+n_{(-)}}=(-1)^{p(n-p)+n_{(-)}}=(-1)^{(n+1)p+n_{(-)}}$[this is so because $p(p+1)=even$ $number$], for
$n$ odd the factor in (\ref{2.14}) is independent on $p$; thus the distinction between $*$ and $\star$ becames 
essential only when $n$ = even.
The definition (\ref{2.11}) specialized for $\alpha = 1 \in\Lambda^0$ gives
\begin{equation}
\label{2.16}
  *1 = \mid det(g_{\alpha\beta}) \mid^{\frac{1}{2}}dx^1 \wedge ...\wedge dx^n = 
  \mid det(g_{ab}) \mid^{\frac{1}{2}} e^1 \wedge ... \wedge e^n;
\end{equation}
consequently, $*1$ can be interpreted as the invariant element of volume 
on $V_n$. Of course, $*$ is a linear operation, $*(\alpha + \beta) = *\alpha + *\beta$. Moreover,
if $deg \alpha = 0, \,\, *(\alpha \wedge \beta) = \alpha\wedge*\beta$.
Thus $*\alpha$ when $\alpha \in \Lambda^0$ can be interpreted as $\alpha\cdot*1$.
Now, (\ref{2.13}) specialized for $q = 0$ gives
\begin{eqnarray}
\label{2.17}
deg(\alpha) = deg(\beta) = p \rightarrow  *(\alpha \wedge *\beta) & = & *(\beta \wedge *\alpha)\\ 
 & = & (-1)^{n_{(-)}}p!\alpha_{a_1...a_p}\beta^{a_1...a_p} \nonumber.
\end{eqnarray}
Applying here (\ref{2.14})
\begin{equation}
\label{2.18a}
 deg(\alpha) = deg(\beta) = p \rightarrow \alpha \wedge *\beta = \beta \wedge *\alpha =
 p!\,\alpha_{a_1...a_p}\beta^{a_1...a_p}*1.
\end{equation}
The same rewritten in terms of forms normalized as in (\ref{2.5}) amounts to:
\begin{equation}
\label{2.18b}
  deg\buildrel{\sim}\over{\alpha} = deg\buildrel{\sim}\over{\beta} = p \,\,\, \rightarrow \,\,\, \buildrel{\sim}\over{\alpha}
  \wedge *\buildrel{\sim}\over{\beta} = \buildrel{\sim}\over{\beta} \wedge *\buildrel{\sim}\over{\alpha}=
  \frac{1}{p!}\alpha_{a_1 ... a_p}\beta^{a_1 ... a_p} *1.
\end{equation}
Thus, if $n_{(-)} = 0$ and the metric is positive definite, $\alpha \wedge *\alpha$
is proportional to $*1$ with the non-negative proportionality coefficient 
which vanishes only when $\alpha = 0$; this remark will be important later.

If $\alpha,\beta \in \Lambda^p$ are simple {\it i.e.}, when
\begin{equation}
\label{2.19a}
 \alpha = \buildrel{1}\over{\alpha} \wedge ... \wedge \buildrel{p}\over{\alpha},\,\, 
 \beta = \buildrel{1}\over{\beta} \wedge ... \wedge \buildrel{p}\over{\beta} \,\, {\rm where} \,
 \buildrel{i}\over{\alpha}={\buildrel{i}\over{\alpha}}_\mu dx^\mu ,~~
 \buildrel{i}\over{\beta}={\buildrel{i}\over{\beta}}_\mu dx^\mu,
\end{equation}
then denoting $(i,j)$ the scalar product $g^{\mu\nu}\alpha_\mu\beta_\nu$ we easily 
find that
\begin{equation}
\label{2.19b}
     \alpha \wedge *\beta = \beta \wedge * \alpha = det(i,j)*1 \,\,\,\,\, {\rm with} \,\, i,j = 1,...,p.
\end{equation}
Sometimes (\ref{2.19b}) with $*1$ denoting the invariant element of volume serves as the 
starting point when one defines the star operation: the Gramm's determinant $det(i,j)$ (or the
$Grammian$) can be considered as the natural generalization of the scalar product of vectors 
on the level of the simple $p$-forms which are sometimes called multivectors ($p$-vectors).

\subsection{The co-multiplication}

In the case of the metrical geometry where the algebra $\Lambda$ besides the basic operations + and $\wedge$ is equipped
with the map $*$, it is convenient to define some secondary algebraic operations on $\Lambda$ which are useful in
elucidating the dual symmetries in the calculus of forms ~\cite{Ple},~\cite{Pleb2}.

We define the $co-multiplication$ (external co-multiplication) denoted by $\buildrel{*}\over{\wedge}$ 
as the map
\begin{equation}
\label{3.1}
  \buildrel{*}\over{\wedge}: \Lambda^{n-p} \times \Lambda^{n-q} \,\,\, \rightarrow \,\,\,
  \Lambda^{n-p-q},
\end{equation}
defined by the condition: for every $\alpha, \beta$ of fixes degrees
\begin{equation}
\label{3.2a}
 *(\alpha \wedge \beta):= *\beta \buildrel{*}\over{\wedge} *\alpha,
\end{equation}
equivalent to
\begin{equation}
\label{3.2b}
*(\alpha \buildrel{*}\over{\wedge} \beta) = (-1)^{n_{(-)}}(*\beta \wedge *\alpha).   \footnote {Notation of this section, in particular the concept of $\buildrel{*}\over{\wedge}$ composition,
deviates from the standard exposition of the calculus of forms. The material presented here can be equivalently
formulated in terms of alternative notations, e.g., in terms of the sometime used $\rfloor$ (step product) composition of forms. In authors
opinion, however, the notation of this section is more advantageous, hence its use.} 
\end{equation}

The composition $\alpha \buildrel{*}\over{\wedge} \beta$ defined by (\ref{3.2a}) is bilinear; is also associative as the implication of
the associativity of the $\wedge$ composition. It is obvious that $\buildrel{*}\over{\wedge}$ can be also considered as the bilinear
map $\buildrel{*}\over{\wedge} : \Lambda \times \Lambda \,\, \rightarrow \,\, \Lambda; ~ \alpha \buildrel{*}\over{\wedge} \beta$ is well defined for
every $\alpha,\beta \in \Lambda$.
The explicit construction of $\alpha \buildrel{*}\over{\wedge} \beta$ for $\alpha$ and $\beta$ of 
fixed degrees amounts to
\begin{equation}
\label{3.3a}
 \alpha \in \Lambda^{n-p}, \beta \in \Lambda^{n-q} \,\, \rightarrow \,\, \alpha \buildrel{*}\over{\wedge} \beta =
 (-1)^{(p+q)(n-p-q)}*(*\beta \wedge *\alpha) \ni  \Lambda^{n-p-q}.
\end{equation}
One easily sees that the co-multiplication $\buildrel{*}\over{\wedge}$ fairly well imitates the properties of $\wedge$ multiplication.
While (\ref{2.6}) means that $deg~(\alpha \wedge \beta) = deg~\alpha + deg~\beta$, (\ref{3.1}) means that
$codeg~(\alpha \buildrel{*}\over{\wedge}\beta)=codeg~\alpha + codeg~\beta;$ (\ref{2.7}) is parallel to:
\begin{equation}
\label{3.4}
  *1 \buildrel{*}\over{\wedge} \alpha = \alpha \buildrel{*}\over{\wedge} *1 = \alpha,
\end{equation}
valid for every $\alpha \in \Lambda$; thus, $*1$ is the unity of the $\buildrel{*}\over{\wedge}$ composition. 
Also parallely to (\ref{2.8}) which says that when $\alpha, \beta$ have the fixed degrees, $\alpha \wedge \beta =
(-1)^{deg~\alpha \cdot deg~\beta} \beta \wedge \alpha$, we have:
\begin{equation}
\label{3.5}
 \alpha \in \Lambda^{n-p}, \,\, \beta \in \Lambda^{n-q} \,\, \rightarrow \,\, \alpha \buildrel{*}\over{\wedge} \beta = 
 (-1)^{pq} \beta \buildrel{*}\over{\wedge} \alpha,
\end{equation}
or $\alpha \buildrel{*}\over{\wedge} \beta = (-1)^{codeg~\alpha \cdot codeg~\beta} 
\beta \buildrel{*}\over{\wedge} \alpha $.

$\big[$In base of this we can define other operation, $\buildrel{\wedge}\over{*}$ like follows
\begin{equation}
\buildrel{\wedge}\over{*} = (-1)^{codeg~\alpha \cdot codeg~\beta} \buildrel{*}\over{\wedge},
\end{equation}
thus we can easily obtain the relation 
\begin{equation}
\beta \buildrel{*}\over{\wedge} \alpha = \alpha \buildrel{\wedge}\over{*} \beta,
\end{equation}
with the properties
\begin{equation}
* \alpha \buildrel{\wedge}\over{*} * \beta = * \beta \buildrel{*}\over{\wedge} * \alpha = *(\alpha \wedge \beta). \big]
\end{equation}

Finally parallel to (\ref{2.9}) we have:
\begin{equation}
\label{3.6}
  (\alpha \buildrel{*}\over{\wedge} \beta)\buildrel{*}\over{\wedge} \gamma = \alpha \buildrel{*}\over{\wedge}(\beta \buildrel{*}\over{\wedge}
  \gamma),
\end{equation}
or in other words the following diagram 

\begin{equation}
\label{3.6,1}
\begin{array}{ c c c }
   \Lambda \times \Lambda \times \Lambda &\harr{id \times \buildrel{*}\over{\wedge}}{}&\Lambda \times
   \Lambda\\
   \varrow{\buildrel{*}\over{\wedge} \times id}{}& &\varrow{\buildrel{*}\over{\wedge}}{}\\ 
  \Lambda \times \Lambda & \harr{\buildrel{*}\over{\wedge}}{} & \Lambda\\
  \end{array}
\end{equation}
is commutative for every $\alpha,\beta,\gamma$ of definite degrees.

These properties show that the ``co-multiplication" seems to deserve its name. The basic difference of the $\wedge$ and
$\buildrel{*}\over{\wedge}$ compositions is that for $\alpha,\beta$ of definite degrees, $deg(\alpha \wedge \beta)$ is $\geq$ than the degrees
of factors (equal for at least one of the factors in $\Lambda^0$) while $deg(\alpha \buildrel{*}\over{\wedge} \beta) \leq$ than the degrees of
factors (equal for at least one factor in $\Lambda^n$).
Notice that $\alpha \buildrel{*}\over{\wedge} \beta$ can be non-trivial only for $n \geq codeg \alpha + codeg \beta
\geq 0 \longleftrightarrow 2n \geq deg \alpha + deg \beta \geq n$. Parallely to $deg \alpha + deg \beta >n
\rightarrow \alpha \wedge \beta =0$, we have $codeg \alpha + codeg \beta >n \rightarrow \alpha 
\buildrel{*}\over{\wedge} \beta  = 0$.
Now, one easily finds that using the concept of $\buildrel{*}\over{\wedge}$ we can rewrite
(\ref{2.13}) in the form:
\begin{equation}
\label{3.7}
  deg\alpha = p, deg\beta = p+q \,\, \rightarrow \,\, *\alpha \buildrel{*}\over{\wedge} \beta =
  \frac{(p+q)!}{q!} \alpha^{b_1 .. b_p}\beta_{b_1 .. b_pa_1..a_p}e^{a_1} \wedge .. \wedge e^{a_q}, 
\end{equation}
what contains as the special case:
\begin{equation}
\label{3.8}
deg\alpha = deg\beta = p \,\, \rightarrow \,\, *\alpha \buildrel{*}\over{\wedge} \beta = *\beta \buildrel{*}\over{\wedge}
\alpha= p! \alpha_{a_1 .. a_p}\beta^{a_1..a_p} \in \Lambda^0.
\end{equation}
Notice that (\ref{3.7}) is equivalent to:
\begin{eqnarray}
\label{3.9}
&&\alpha \in \Lambda^{n-p}, \beta \in \Lambda^{n-q} \rightarrow \alpha \buildrel{*}\over{\wedge} \beta 
\\&& = \mid
det(g_{ab})\mid^{-\frac{1}{2}} \beta_{a_1 .. a_pb_1..b_{n-p-q}}\epsilon^{a_1..a_pc_1..c_{n-p}} \alpha_{c_1..c_{n-p}}
\cdot (^{n-q}_{\,\,\,p})e^{b_1} \wedge...\wedge e^{b_{n-p-q}}.\nonumber
\end{eqnarray}
This can be interpreted as the explicit form (\ref{3.3a}) in terms of the scalar components of the tensors
defining $\alpha$ and $\beta$.
If (\ref{3.9}) would be considered as the definition of $\alpha \buildrel{*}\over{\wedge} \beta$, the
associativity of the $\buildrel{*}\over{\wedge}$ composition becomes non-trivial property. Notice that
(\ref{3.9}) specialized for $\alpha \in \Lambda^{n-p}, \beta \in \Lambda^{n-(n-p)}=\Lambda^p$ is
\begin{equation}
\label{3.10}
 \alpha \buildrel{*}\over{\wedge} \beta =\mid det(g_{ab})\mid^{-\frac{1}{2}}
 \beta_{a_1...a_p}\epsilon^{a_1...a_pb_1...b_{n-p}}\alpha_{b_1 ...b_{n-p}} \in \Lambda^0,
\end{equation}
is an equivalent form of (\ref{3.8}).

Now, (\ref{2.13}) implies: $(e_a := g_{ab}e^b)$
\begin{equation}
\label{3.11a}
(e_{a_1} \wedge..\wedge e_{a_p}) \wedge *(e^{b_1} \wedge ... \wedge e^{b_{p+q}}) = 
\frac{1}{q!} *(e^{c_1} \wedge ...\wedge e^{c_q}) \delta^{b_1 ..... b_{p+q}}_{c_1..c_qa_1..a_p},
\end{equation}
while (\ref{3.7}) gives a dual form of this identity
\begin{equation}
\label{3.11b}
*(e_{a_1} \wedge..\wedge e_{a_p}) \buildrel{*}\over{\wedge}(e^{b_1} \wedge ... \wedge e^{b_{p+q}}) =
\frac{1}{q!} \delta^{b_1 ..... b_{p+q}}_{a_1..a_pc_1..c_q} e^{c_1} \wedge ...\wedge e^{c_q}.
\end{equation}
In particular, (\ref{3.11a}),~(\ref{3.11b}) imply
\begin{equation}
\label{3.12a}
e^a \wedge *e^b = g^{ab} \cdot *1 \,\, , \,\,\,\,\, *e^a \buildrel{*}\over{\wedge} e^b = g^{ab},
\end{equation}
what can be interpreted as
\begin{equation}
\label{3.12b}
dx^\alpha \wedge *dx^\beta = g^{\alpha \beta} \cdot *1 \,\, , \,\,\,\,\, *dx^\alpha \buildrel{*}\over{\wedge}
dx^\beta = g^{\alpha \beta}.
\end{equation}
We defined $*$ treating the Riemannian metric as given; the relations (\ref{3.12a}),
(\ref{3.12b}) suggest a possibility of an inverse construction: assuming in $\Lambda$ the existence of the formal map $*$ equipped in the adequate abstract properties,
one can considers the Riemannian metric as determined by $g^{\alpha\beta} = *dx^\alpha \buildrel{*}\over{\wedge}
dx^\beta$.\footnote{A further extension of this remark stays beyond the scope of this text.}

The formula (\ref{3.11b}) has some important technical application: if $\Lambda^{p+q} \ni \alpha =
\alpha_{a_1...a_{p+q}}e^a_1 \wedge ...\wedge e^a_{p+q}$, then according to (\ref{3.11b}) we have
\begin{equation}
\label{3.13}
*(e_{a_1} \wedge..\wedge e_{a_p}) \buildrel{*}\over{\wedge} \alpha = \frac{(p+q)!}{q!}
\alpha_{a_1...a_pb_1...b_q}e^{b_1} \wedge ...\wedge e^{b_q}.
\end{equation}
Therefore, the operation $*(e_{a_1}\wedge ... \wedge e_{a_p}) \buildrel{*}\over{\wedge}...$ determines a process of
peelling~of the forms $e^a$ from the arbitrary form $\alpha$.
This process will be very useful later, when we will discuss the structure equations.
The composition $\buildrel{*}\over{\wedge}$ is in general not associative $with \,\, respect$ to the $\wedge$ composition:
in general $(\alpha \buildrel{*}\over{\wedge} \beta)\wedge \gamma \neq \alpha \buildrel{*}\over{\wedge} (\beta \wedge
\gamma)$ [the degrees of both sides are the same].
One easily shows, however that (\ref{3.7}) implies
\begin{equation}
\label{3.14a}
 deg \gamma = 1 \rightarrow *\gamma \buildrel{*}\over{\wedge}(\alpha \wedge \beta) = (*\gamma \buildrel{*}\over{\wedge}
 \alpha)\wedge \beta +(-1)^{deg\alpha} \alpha \wedge(*\gamma \buildrel{*}\over{\wedge} \beta). 
\end{equation}
By application of $*$ on (\ref{3.14a}) an appropriate changes of symbols one easily obtains
\begin{equation}
\label{3.14b}
deg \gamma = 1 \rightarrow \gamma \wedge(\alpha \buildrel{*}\over{\wedge} \beta) = (\gamma \wedge
 \alpha)\buildrel{*}\over{\wedge} \beta +(-1)^{codeg\alpha} \alpha \buildrel{*}\over{\wedge}(\gamma \wedge \beta). 
\end{equation}

\subsection{Algebraic operations on forms}

The formulas (\ref{3.14a}),~(\ref{3.14b}) have the structure of the Leibnitz rules, they suggest that $*\gamma
\buildrel{*}\over{\wedge}...$ acts on $\alpha \wedge \beta$ like a sort of the algebraic differentiation, while $\gamma
\wedge ...$ acts on $(\alpha \buildrel{*}\over{\wedge} \beta)$ in a similar manner; the same can be stated precisely as
follows:

Let  $\gamma, \gamma' \in \Lambda^1$ and let $m[\gamma], \mu[\gamma']$ denote two maps:
\begin{equation}
\label{3.15a}
m[\gamma]: \Lambda^p \rightarrow \Lambda^{p+1},~m[\gamma]\alpha = \gamma \wedge \alpha, \,\,\, deg\alpha =p,
\end{equation}
\begin{equation}
\label{3.15b}
\mu[\gamma']: \Lambda^p \rightarrow \Lambda^{p-1}, ~\mu[\gamma']\alpha = *\gamma' \buildrel{*}\over{\wedge} \alpha, 
\,\,\, deg\alpha=p.
\end{equation}
Then, with arbitrary $\gamma,\gamma'$ we have
\begin{equation}
\label{3.16a}
m(\alpha \wedge \beta) = m\alpha \wedge \beta + (-1)^{deg\alpha}\alpha \wedge m\beta,
\end{equation}
\begin{equation}
\label{3.16b}
m(\alpha \buildrel{*}\over{\wedge} \beta) = m\alpha \buildrel{*}\over{\wedge} \beta + 
(-1)^{codeg\alpha}\alpha \buildrel{*}\over{\wedge} m\beta,
\end{equation}
and
\begin{equation}
\label{3.17a}
\mu(\alpha \wedge \beta) = \mu \alpha \wedge \beta + (-1)^{deg\alpha}\alpha \wedge \mu \beta,
\end{equation}
\begin{equation}
\label{3.17b}
\mu (\alpha \buildrel{*}\over{\wedge} \beta) = \mu \alpha \buildrel{*}\over{\wedge} \beta +
(-1)^{codeg\alpha}\alpha \buildrel{*}\over{\wedge} \mu \beta.
\end{equation}
The maps $m$ and $\mu$ fulfill the anti-commutation rules
\begin{equation}
\label{3.18a}
m[\gamma]m[\gamma'] + m[\gamma']m[\gamma] = 0,
\end{equation}
\begin{equation}
\label{3.18b}
\mu[\gamma] \mu[\gamma'] + \mu[\gamma'] \mu[\gamma] = 0,
\end{equation}
\begin{equation}
\label{3.18c}
m[\gamma] \mu[\gamma'] + \mu[\gamma']m[\gamma] = (*\gamma \buildrel{*}\over{\wedge} \gamma')id.
\end{equation}
Of course, $*\gamma \buildrel{*}\over{\wedge} \gamma' = \gamma_a \gamma'^a$; $id$ denotes the identity map. [rules
(\ref{3.18a}), (\ref{3.18b}) follow from the skewness of $\gamma \wedge \gamma'$ and $*\gamma \buildrel{*}\over{\wedge} *\gamma'$;
(\ref{3.18c}) one proves using (\ref{3.16b}), (\ref{3.17a}).] The rules (\ref{3.18a}), (\ref{3.18b}),
(\ref{3.18c}) are useful in the demonstration of the isomorphism of the
algebra $\Lambda$ and the $2^n$ Clifford algebra.
In particular, (\ref{3.18a}), (\ref{3.18b}) with $\gamma = \gamma'$ imply that the both discussed 
maps are nilpotent
\begin{equation}
\label{3.19}
m(m\alpha) = 0, \,\,\,\,\, \mu(\mu \alpha) = 0.
\end{equation}
We will say that a form $\alpha$ is $\gamma-closed$ if $m[\gamma]\alpha=0$ and $*\gamma-closed$ when
$\mu[\gamma]\alpha=0$ (of course, $\gamma \in \Lambda^1$). 
An useful lemma easily follows:
\begin{equation}
\label{3.20a}
m[\gamma]\alpha = 0 \,\,\, \leftrightarrow \,\,\, {\rm there \,\, exist\,\, such\,\, a\,\, form} \,\,
\beta \,\, {\rm that}\,\, \alpha = m[\gamma]\beta,
\end{equation}
\begin{equation}
\label{3.20b}
\mu[\gamma]\alpha = 0 \,\,\, \leftrightarrow \,\,\, {\rm there \,\, exist\,\, such\,\, a\,\, form}\,\,
 \beta\,\, {\rm that}\,\, \alpha = \mu[\gamma]\beta.
\end{equation}
Due to (\ref{3.19}) the implications $\leftarrow$ are obvious; the proof of the inverse implication is contained in
(\ref{3.18c}): if $m[\gamma]\alpha =0$, one can so select $\gamma'$ that $\gamma_a {\gamma'}^a \neq 0$; therefore
(\ref{3.18c}) gives $\alpha = m[\gamma]\mu[\gamma'](\gamma_a {\gamma'}^a)^{-1}\alpha$; similarly, when $\mu[\gamma']\alpha =0$,
one can so select $\gamma$ that $\gamma_a {\gamma'}^a \neq 0$; then (\ref{3.18c}) yields: $\alpha = \mu[\gamma']m[\gamma](\gamma_a
{\gamma'}^a)^{-1}\alpha$.
We can observe that due to (\ref{3.16b}) and (\ref{3.17a}) and the nilpotence of the maps $m, \mu$, the set of $\gamma-closed$
forms is closed with respect to the $\buildrel{*}\over{\wedge}$, while the set of $*\gamma-closed$ forms is closed
with respect to the $\wedge$ composition
\begin{equation}
\label{3.21a}
m\alpha \buildrel{*}\over{\wedge} m\beta = m(\alpha \buildrel{*}\over{\wedge} m\beta), 
\end{equation}
\begin{equation}
\label{3.21b}
\mu\alpha \wedge \mu\beta = \mu(\alpha \wedge \mu\beta).
\end{equation}
(This is accompanied by the obvious $m \alpha \wedge m\beta =0,~\mu\alpha \buildrel{*}\over{\wedge} \mu\beta =0$; of
course, $m's$ and $\mu's$ with suppresed argument depend on a fixed $\gamma \in \Lambda^1).$
We will close these considerations noticing that for every $\alpha \in \Lambda^p$ and 
$\gamma \in \Lambda^1$
\begin{equation}
\label{3.22a}
m[\gamma]\alpha = (-1)^{np-n_{(+)}+1}*\mu[\gamma]*\alpha,
\end{equation}
\begin{equation}
\label{3.22b}
\mu[\gamma]\alpha = (-1)^{np-n_{(+)}}*m[\gamma]*\alpha.
\end{equation}
Further on, we will see that the properties of the algebraic maps $m$ and $\mu$ are very similar to the properties of
the basic differential maps $d$ and $\delta$.

\begin{center}
\section{Differential operations on forms}
\end{center}
\setcounter{equation}{0}

The analytic theory of forms uses the two basic differential equations: the external differential $d$ and the
external co-differential $\delta$. The $d$ operation generalizes the concepts of the gradient and of the rotation; the
$\delta$ operation generalizes the concept of the divergence. This section describes the basic properties of the $d$
operation ~\cite{GS},~\cite{Flan},~\cite{Schutz},~\cite{Nak}.

\null

\subsection{The external differential}
The operation of the {\it external differential d} is a map
\begin{equation}
\label{4.1}
d: \Lambda^p \,\,\, \rightarrow \,\,\, \Lambda^{p+1},
\end{equation}
defined for $\alpha \in \Lambda^p$ with the local representation (\ref{2.4}) by the formula
\begin{equation}
\label{4.2}
d\alpha = \alpha_{\mu_1 ... \mu_p,\lambda}dx^\lambda \wedge dx^{\mu_1} \wedge ... \wedge dx^{\mu_p} \,\,\, \in
\Lambda^{p+1}.
\end{equation}
Is important to note, that $d\alpha$ does not depend on any local map and represents a global 
concept; (\ref{4.2}) implies:
\begin{equation}
\label{4.3}
 {\rm for \,\, every} \,\, \alpha: d(d\alpha)=0.
\end{equation}
This relation interpreted as a global relation, states that the map $d$ is nilpotent and is 
called the {\it Poincar\'e's ~Lemma} ~\cite{Flan}. 

One easily shows that $d$ fulfills the $Leibnitz ~ rule$
\begin{equation}
\label{4.4}
d(\alpha \wedge \beta) = d\alpha \wedge \beta + (-1)^{deg\alpha} \alpha \wedge d\beta.
\end{equation} 
A form $\alpha$ such that $d\alpha=0$ is called closed (or homological to zero). The differential of any
differentiated form is closed, but not always a closed form can be represented (globally) as a differential of some other form. At this
point the global topological structure of $M_n$ is very essential. It is valid, however, the local inversion of the
Poincar\'e's lemma which states that $d\alpha =0$ true in a singly-connected region of $M_n$ guarantees the existence
in this region of such a form $\beta$ that $\alpha = d\beta$. The set of differentials is closed with respect to the
composition $\wedge$, due to (\ref{4.4}) and (\ref{4.3}) $d(d\alpha \wedge d\beta) = 0$, moreover
\begin{equation}
\label{4.5}
 d\alpha \wedge d\beta = d(\alpha \wedge d\beta).
\end{equation} 
Of course, we can consider $d$ as the map $d:\Lambda \rightarrow \Lambda$ defined by the partial maps: $d\alpha =
\sum^n_{(p=0)}\,\, d\alpha_p$.

The basic definition (\ref{4.2}) implies that $d\alpha$ can be represented in the form $(\alpha \in \Lambda^p)$
\begin{eqnarray}
\label{4.6}
d\alpha & = & (-1)^p \alpha_{\mu_1 ... \mu_p;\mu_{p+1}}dx^{\mu_1}\wedge...\wedge dx^{\mu_{p+1}} \\ & = &
[(-1)^p \alpha_{a_1...a_p,a_{p+1}}+ p \alpha_{r a_1...a_{p-1}}\Gamma^r_{a_pa_{p+1}}]e^{a_1}\wedge...\wedge e^{a_{p+1}}
\nonumber \\ & = &
(-1)^p \alpha_{a_1...a_p,a_{p+1}}e^{a_1}\wedge...\wedge e^{a_{p+1}} + \alpha_{a_1...a_p}d(e^{a_1}\wedge...\wedge
e^{a_p}).
\nonumber
\end{eqnarray}
The second line presents $d\alpha$ as defined independently of the choice of the local maps on $M_n$. The
differential in the third line can be written as a compact expression
\begin{eqnarray}
\label{4.7}
d(e^{a_1}\wedge ... \wedge e^{a_p})  & = & \frac{(-1)^{p-1}}{(p-1)!} \delta^{a_1...a_p}_{b_1...b_{p-1}r}
\Gamma^r_{b_pb_{p+1}}e^{b_1} \wedge ... \wedge e^{b_{p+1}} \nonumber \\  & = & \nonumber
\frac{(-1)^{p+1}}{(p+1)!} \left [ \delta^{a_1...a_{p+2}}_{b_1...b_{p+2}}\Gamma^{b_{p+2}}_{a_{p+1}a_{p+2}}
\right. \nonumber \\
& & \left. \hspace{3em} 
- \delta^{a_1...a_{p+1}}_{b_1...b_{p+1}}\Gamma^{r}_{a_{p+1}r} \right ] e^{b_1} \wedge ... \wedge e^{b_{p+1}}.
\end{eqnarray}
In particular, we have the obvious $d(e^{a_1} \wedge ... \wedge e^{a_n})=0$, 
\begin{eqnarray}
\label{4.8}
d(e^{a_1} \wedge ... \wedge e^{a_{n-1}}) & = &(-1)^n e^{a_1} \wedge ... \wedge e^{a_n} {{\Gamma_{a_n}}^r}_r
 \nonumber \\ & = &  * \left [ (-1)^n {\mid det(g_{ab})\mid}^{-\frac{1}{2}} 
 \epsilon^{a_1 ... a_n}{{\Gamma_{a_n}}^r}_r \right ].  
\end{eqnarray}
Thus, the external differential $d$ induces the sequence:
\begin{equation}
\label{4.8,1}
0 \harr{i}{} \Lambda^0 \harr{d}{} \Lambda^1 \harr{d}{} ... \harr{d}{} \Lambda^{n-1} \harr{d}{}
\Lambda^n \harr{d}{} 0,
\end{equation}
where $i$ is the inclusion map.
 
The above sequence is called {\it de Rham complex}, and since $d^2=0$ we have 
that $im(d) \subset ker(d)$.

\subsection{Structure equations}
On the other extreme, (\ref{4.7}) specialized for $p=1$ gives
\begin{equation}
\label{4.9}
 de^a = {\Gamma^a}_{bc}e^b \wedge e^c.
\end{equation}
This using the concept of the Ricci forms introduced in the first section can be rewritten as
\begin{equation}
\label{4.10a}
 de^a = e^b \wedge {\Gamma^a}_b.
\end{equation}
This is the first of the Cartan $structure ~ equations$ \cite{Chan},~\cite{Pleb1},~\cite{Torres},~\cite{Choquet}, in the case 
of zero torsion; when ${T^\alpha}_{\beta \gamma} \neq 0$ the first structure equation is given by
\begin{equation}
\label{4.101}
de^a + {\Gamma^a}_b \wedge e^b = T^a,
\end{equation}
where $T^a = \frac{1}{2} {T^a}_{bc} e^b \wedge e^c$ the {\it torsion two-form}.

Now (\ref{4.10a}) supplemented by the condition $\Gamma^{ab} = \Gamma^{[ab]}$ determines 
uniquely ${\Gamma^a}_b$ when $e^a$ is treated as given. The second structure 
equation is:
\begin{equation}
\label{4.10b}
d{\Gamma^a}_b + {\Gamma^a}_s \wedge {\Gamma^s}_b = \frac{1}{2}{R^a}_{bcd}e^c \wedge e^d :=
\frac{1}{2}R^a_b,
\end{equation}
where $R^{ab} = R^{[ab]} \in \Lambda^2$ is called the curvature form.
Equations (\ref{4.10b}) easily follow from the original definition of ${\Gamma^a}_b$ with $;\nu$ derivative by the
application of the standard Ricci formula for the commutator of the covariant derivatives.
The Poincar\'e's lemma implies the necessity of the integrability conditions of the structure equations: the
consistence of (\ref{4.10a}) with $d(de^a)=0$ amounts to
\begin{equation}
\label{4.11a}
e^b \wedge R^a_b =0,
\end{equation}
equivalent to $R^a_{[bcd]}=0$, while the consistence of (\ref{4.10b}) with $d(d{\Gamma^a}_b) = 0$ amounts to:
\begin{equation}
\label{4.11b}
dR^a_b = R^a_s \wedge {\Gamma^s}_b - {\Gamma^a}_s \wedge R^s_b.
\end{equation}
The last equation is equivalent to $R^a_{b[cd;e]} =0$, i.e., to the Bianchi identities written 
in terms of the scalar components.
Notice that (\ref{4.11a}),(\ref{4.11b}) already close the chain of the integrability conditions. When 
(\ref{4.10a}), (\ref{4.10b}) and (\ref{4.11a}), (\ref{4.11b}) are assumed, then applying $d$ 
on (\ref{4.11a}), (\ref{4.11b}) according to the general rules of the game, one obtains only 
identities 0 = 0.

\subsection{The co-differential}

The external co-differential $\delta$ is a map
\begin{equation}
\label{5.1}
 \delta: \,\,\, \Lambda^p \,\, \rightarrow \,\, \Lambda^{p-1},
\end{equation}
defined as a composition of the maps previously discussed ~\cite{Ple},~\cite{Pleb2}
\begin{equation}
\label{5.2}
deg\alpha = p \rightarrow \delta\alpha = (-1)^{np-n_{(+)}+1}*d*\alpha.
\end{equation}
The normalization factor is so selected that when $deg \alpha = p,~deg\beta =p+1 \rightarrow
deg(\alpha \wedge *\beta)=n-1,~deg (d(\alpha \wedge *\beta))=n, \,\,$we have
\begin{equation}
\label{5.3}
d(\alpha \wedge *\beta) = d\alpha \wedge *\beta - \alpha \wedge *\delta \beta.
\end{equation}
Notice that (\ref{5.2}) implies as an inverse formula
\begin{equation}
\label{5.4}
deg \alpha = p \rightarrow d\alpha =(-1)^{np-n_{(+)}} *\delta *\alpha.
\end{equation}
We have also
\begin{equation}
\label{5.5}
deg\alpha = p \rightarrow *d\alpha = (-1)^{p+1} \delta *\alpha, ~~ *\delta \alpha = (-1)^p d*\alpha.
\end{equation}
Now, all basic properties of the $d$ operation posses the corresponding co-images in terms of the $\delta$
operation. 
First, the $\delta$ operation is nilpotent
\begin{equation}
\label{5.6}
{\rm for \,\, every} \  \alpha: \delta(\delta \alpha) = 0.
\end{equation}
This is the co-image of the Poincar\'e's lemma. The co-image of the Leibnitz rule (\ref{4.4}) amounts to
\begin{equation}
\label{5.7}
 \delta(\alpha \buildrel{*}\over{\wedge} \beta) = \delta \alpha \buildrel{*}\over{\wedge} \beta +
 (-1)^{codeg \alpha} \alpha \buildrel{*}\over{\wedge} \delta \beta,
\end{equation}
$i.e.$, it is a Leibnitz rule but with respect to the co-multiplication.

A form $\alpha$ such that $\delta \alpha = 0$ is called co-closed (or co-homological to zero). The
co-differential of any co-differentiated form is closed. But not every co-closed form can be represented (globally) as a
co-differential of some other form. At this point the global structure of $M_n$ is essential. The dual variant
of the local inversion of the Poincar\'e's lemma assures, however that if $\delta \alpha =0$ in a singly connected
region of $M_n$, then $\alpha$ can be represented as $\alpha = \delta \beta$ in this region.
Then, exactly like with (\ref{4.8,1}) we have that the external codifferential $\delta$ induces the
sequence
\begin{equation}
0 \harrl{\delta}{} \Lambda^0 \harrl{\delta}{} \Lambda^1 \harrl{\delta}{} ... \harrl{\delta}{} 
\Lambda^{n-1} \harrl{\delta}{} \Lambda^n \harrl{i}{} 0
\end{equation}
where $i$ is the inclusion map, and since ${\delta}^2=0$ we have that $im(\delta) \subset ker(\delta)$.

The set of co-differentials is closed with respect to the comultiplication. Indeed, due 
to (\ref{5.6}), (\ref{5.7}) $\delta(\delta \alpha \buildrel{*}\over{\wedge} \delta \beta) =0$, moreover
\begin{equation}
\label{5.8}
 \delta \alpha \buildrel{*}\over{\wedge} \delta \beta = \delta (\alpha \buildrel{*}\over{\wedge} 
  \delta \beta).
\end{equation}
Of course, $\delta$ can be also interpreted as the map $\delta : \Lambda \rightarrow \Lambda$, defined in terms
of the ``partial" maps according to $\delta \alpha = \sum^n_{p=0} \,\, \delta \alpha_p$.
Now, the local representation of $\delta \alpha,~deg~\alpha =p$ amounts to
\begin{equation}
\label{5.9}
\delta \alpha = (-1)^p p {\alpha_{\mu_1...\mu_{p-1}\lambda}}^{;\lambda} 
dx^{\mu_1}\wedge .... \wedge dx^{\mu_{p-1}}.
\end{equation}
This can be also rewritten in terms of the scalar components, as an expression insensitive on the 
choice of the local maps
\begin{eqnarray}
\label{5.10}
\delta \alpha & = & (-1)^p \left [ p({\alpha_{a_1 ... a_{p-1}s}}^{,s} + \alpha_{a_1 ... a_{p-1}s} {\Gamma_r}^{sr}) +
\right.\\ & & \left. \qquad \ \ p(p-1)\alpha_{a_1 ... a_{p-2}sr} {\Gamma_{a_{p-1}}}^{sr}
\right ]e^{a_1} \wedge ... \wedge e^{a_{p-1}}\nonumber \\ & = & \nonumber
(-1)^p p {\alpha_{a_1 ... a_{p-1}s}}^{,s} e^{a_1} \wedge ... \wedge e^{a_{p-1}}+ 
\alpha_{a_1 ... a_{p}} \delta (e^{a_1} \wedge ... \wedge e^{a_p}),
\end{eqnarray} 
where (for $p \geq 1$)
\begin{equation}
\label{5.11}
\delta(e^{a_1} \wedge ... \wedge e^{a_p})= \frac{(-1)^p}{(p-1)!} \delta^{a_1 ... a_{p+1}}_{b_1 ... b_{p+1}}
{\Gamma_{a_{p+1}}}^{b_pb_{p+1}}e^{b_1} \wedge ... \wedge e^{b_{p-1}}.
\end{equation}
[Of course, ${T_{...}}^{,a}$ denotes $T_{...,b}g^{ba}$.] This general formula contains as special cases
$\delta (e^{a_1} \wedge ... \wedge e^{a_n})=0$ [equivalent to $\delta *1 = (-1)^{n+1}*d1 = 0$] and
\begin{equation}
\label{5.12}
\delta(e^{a_1} \wedge ... \wedge e^{a_{n-1}})= \frac{(-1)^{n-1}}{(n-2)!}\epsilon^{a_1 ... a_n} \epsilon_{b_1 ... b_n}
{\Gamma_{a_n}}^{b_{n-1}b_n} e^{b_1} \wedge ... \wedge e^{b_{n-2}}.
\end{equation}
On the other side of the possible extremal values of $p$ we have
\begin{equation}
\label{5.13}
\delta e^a = -2{\Gamma_b}^{[ab]} = {\Gamma^{ab}}_b = -g^{ab}{{e_b}^{\mu}}_{;\mu} = -g^{ab}e^{-1}(e{e_b}^{\mu})_{,\mu}.
\end{equation}
As far as the Ricci forms are concerned one easily finds
\begin{equation}
\label{5.14}
\delta {\Gamma^a}_b = - {{\Gamma^a}_{bs}}^{,s} + {\Gamma^a}_{bs} {\Gamma^{sr}}_r = ({e^a}_{\mu ; \nu} {e_b}^{\mu})^{;\nu}
\in \Lambda^0.   
\end{equation}
The object $\delta {\Gamma^a}_b$ is very sensitive on the choice of ${e^a} 's$. Indeed, executing the permissible
transformation (\ref{1.5}) which leaves $ds^2$ invariant, one obtains
\begin{equation}
\label{5.15}
\delta \Gamma^{a'b'} = {L^{a'}}_a {L^{b'}}_b \delta \Gamma^{ab} + g^{rs}({L^{a'}}_{r,\mu} {L^{b'}}_s)^{;\mu} - \Gamma^{abc}
({L^{a'}}_a {L^{b'}}_b)_{,c}.
\end{equation}
Thus, by an appropriate choice of the $(^n_2)$ parameters of $O(n_{(+)},n_{(-)})$ [understood as functions on
$M_n$] we can assign to $(^n_2)$ of functions $\delta \Gamma^{ab} = \delta \Gamma^{[ab]}$ any arbitrary values. In particular, one can also select
the $``O_n \  gauge"$ that $\delta {\Gamma^a}_b = 0$. Then at least locally, ${\Gamma^a}_b$
would posses the representation: ${\Gamma^a}_b = \delta {\Omega^a}_b,~{\Omega^a}_b \in \Lambda^2$.

\subsection{The co-structure equations} 
Now, the structure equations (\ref{4.10a}), (\ref{4.10b}) and their integrability conditions 
(\ref{4.11b}) can be equivalently rewritten in the form of the relations which result by the 
application of the star operation; we will call these the co-structure equations and the 
co-integrability conditions. Using the concepts of codifferential $\delta$ and the
co-multiplication $\buildrel{*}\over{\wedge}$ one easily finds that the discussed co-equations can be obtained
formally from (\ref{4.10a}), (\ref{4.10b}) and (\ref{4.11a}), (\ref{4.11b}) when one replaces in 
them the particular symbols according to the scheme
\begin{equation}
\label{5.16}
 d \rightarrow \delta, ~~ \wedge \rightarrow \buildrel{*}\over{\wedge}, ~~ e^a \rightarrow *e^a, 
 ~~ {\Gamma^a}_b \rightarrow -*{\Gamma^a}_b, ~~ R^a_b \rightarrow -*R^a_b.
\end{equation}
Therefore, the co-structure equations and the co-integrability conditions amo\-unts to
\begin{equation}
\label{5.17a}
\delta(*e^a) = *e^b \buildrel{*}\over{\wedge} (-*{\Gamma^a}_b),
\end{equation}
\begin{equation}
\label{5.17b}
\delta(-*{\Gamma^a}_b) +  (-*{\Gamma^a}_s)\buildrel{*}\over{\wedge}(-*{\Gamma^s}_b) = \frac{1}{2}(-*R^a_b),
\end{equation}
and
\begin{equation}
\label{5.18a}
*e^b \buildrel{*}\over{\wedge} (-*R^a_b) = 0,
\end{equation}
\begin{equation}
\label{5.18b}
\delta (-*R^a_b) = (-*R^a_s) \buildrel{*}\over{\wedge} (-* {\Gamma^s}_b) - (-*{\Gamma^a}_s) \buildrel{*}\over{\wedge}
(-*R^s_b).
\end{equation}
Notice that
\begin{equation}
\label{5.19}
-*R^a_b = R^a_{bcd}(*e^c) \buildrel{*}\over{\wedge} (*e^d).
\end{equation}
We will see later that the structure and the equivalent co-structure equations considered together form a set of
relations which permits to study the role of the particular irreducible parts of the curvature in the Riemannian
geometry.

\begin{center}
\section{Hopf algebra and differential Hopf algebra}
\end{center}
\setcounter{equation}{0}

We begin by defining some of the basical concepts that we will need to indicate how $\Lambda$,
the space of differential forms, admits the structure of Hopf algebra with
the previous operations and also the action of a diffferential in some
cases ~\cite{Milnor} -~\cite{Kass}.

First, let $K$ be a ring with identity, then a $K$-module $G$ is an additive abelian group together 
with a function $f: K \times G \rightarrow G~~f(k,g) = kg,$ such that
\begin{eqnarray}
(k+k')g = kg + k'g, &\qquad & (kk')g = k(k'g), \\
k(g + g') = kg +kg', & & 1g = g.
\end{eqnarray}
Now, a {\it graded} $K$-module is a family of $K$-modules $\{ G_n \}$ where the index $n$ runs through the non-negative integers (it is not the direct sum of them). \\
Then for two graded $K$-modules $L$ and $M$, a homomorphism of graded modules $f: L \rightarrow M$ of 
degree $r$ is a family $f=\{f_n:L_n \rightarrow M_{n+r};~n \in Z\}$ of $K$-module homomorphisms $f_n$. 
Thus the composition of homomorphisms of degrees $r$ and $r'$ has degree $r+r'$.

It is important to note that the tensor product of two graded $K$-modules $L$ and $M$ is the 
graded $K$-module given by
\begin{equation}
(L \otimes M)_n = \sum_{p+q=n} L_p \otimes M_q,  \footnote{The tensorial product between $K$-modules
and graded $K$-modules must be denoted by ${\otimes}_K$ indicating over which ring we are working, 
but in brief we only write $\otimes$.}
\end{equation}
and therefore the grading in the tensor product is defined by $deg(l \otimes m) = deg~l +
deg~m$, with $l \in L$ and $m \in M$.

A graded $K$-algebra $V$ is a graded $K$-module endowed with two $K$-module homomorphisms
$\varphi:V \otimes V \rightarrow V$ and $\eta:K \rightarrow V$ each of degree 0, (called the
{\it product} and the {\it unit} respectively), which render commutative
the diagrams

\begin{equation}
\begin{array}{ c c c }
   V \otimes V \otimes V &\harr{\varphi \otimes id}{}&V \otimes V\\
   \varrow{id \otimes \varphi}{}& &\varrow{\varphi}{}\\ 
  V \otimes V & \harr{\varphi}{} & V\\
  \end{array}
\end{equation}  

\begin{equation}
\begin{array}{ c c c c c }
   K \otimes V & \cong & V & \cong & V \otimes K\\
   \varrow{\eta \otimes id}{}& & \cong & & \varrow{id \otimes \eta}{}\\ 
  V \otimes V & \harr{\varphi}{} & V & \harrl{\varphi}{}& V \otimes V\\
  \end{array}
\end{equation} 
 
The $K$-algebra $V$ is commutative if, in addition, it satisfies the axiom 
\begin{equation}
\begin{array}{ c c c }
V \otimes V & \harr{\tau_{V,V}}{} & V \otimes V\\
\\
\varphi ~ \searrow  & &  \swarrow ~ \varphi \\
& V & \\
\end{array}
\end{equation}
where $\tau_{V,V}$ is the flip map:~$\tau_{V,V}(v \otimes v') = (-1)^{deg~v \cdot deg~v'} v' \otimes v$.
In the graded case we require this relation be satisfied, and when we
restrict to the elements of degree zero we recover the usual commutative
law, (compare with ~\cite{Milnor},~\cite{Tak},~\cite{Swee}). 

A graded $K$-coalgebra $U$ over the ring $K$ is a graded $K$-module $U$ 
with two homomorphisms $\Psi ~\footnote{In the literature one uses $\Delta$ instead of $\Psi$ but we reserve this former for the harmonic operator.}: U \rightarrow U \otimes U$ and $\epsilon: U \rightarrow K$ of 
graded $K$-modules (the {\it coproduct} and the {\it counit}), each of
degree 0, such that the diagrams

\begin{equation}
\label{}
\begin{array}{ c c c }
   U &\harr{\Psi}{}&U \otimes U\\
   \varrow{\Psi}{}& &\varrow{id \otimes \Psi}{}\\ 
  U \otimes U & \harr{\Psi \otimes id}{} & U \otimes U \otimes U \\
  \end{array}
\end{equation}  

\begin{equation}
\label{7.7}
\begin{array}{ c c c c c }
   U \otimes U & \harrl{\Psi}{} & U & \harr{\Psi}{} & U \otimes U\\
   \varrow{\epsilon \otimes id}{}& & \cong & & \varrow{id \otimes \epsilon}{}\\ 
  K \otimes U & \cong & U & \cong & U \otimes K\\
  \end{array}
\end{equation}  
are commutative.

If, furthermore, the diagram
\begin{equation}
\begin{array}{ c c c }
 & U & \\
\Psi ~ \swarrow & & \searrow ~\Psi \\
\\
U \otimes U & \harr{\tau_{U,U}}{} & U \otimes U \\
\end{array}
\end{equation}
commutes, where $\tau_{U,U}$ is the same flip map, we say that the
$K$-coalgebra $U$ is cocommutative. 

\null     

The tensor product of two graded $K$-algebras $\Omega$ and $\Sigma$ is their tensor product 
$\Omega \otimes \Sigma$, as graded modules, and form an algebra whose product map is
defined as the composite
 
\begin{equation}
(\Omega \otimes \Sigma) \otimes (\Omega \otimes \Sigma) \harr{id \otimes \tau \otimes id}{}
\Omega \otimes \Omega \otimes \Sigma \otimes \Sigma \harr{\varphi_\Omega 
\otimes \varphi_\Sigma}{}
\Omega \otimes \Sigma,
\end{equation} 

\noindent where $\tau$ is the transposition map $\tau[l \otimes m] = (-1)^{deg~l \cdot deg~m} m \otimes l$, and
with unit element map given by $I_\Omega \otimes I_\Sigma:K \cong K \otimes K
\rightarrow \Omega \otimes \Sigma$.
In similar way if $W$ and $W'$ are graded coalgebras, their tensor product $W \otimes W'$ (as
graded modules) is a graded coalgebra with diagonal map the composite

\begin{equation}
W \otimes W' \harr{\Psi \otimes \Psi '}{} W \otimes W \otimes W' \otimes W' \harr{id \otimes
\tau \otimes id}{} (W \otimes W') \otimes (W \otimes W'),
\end{equation}

\noindent $\tau$ the same transposition map, and with counit $\epsilon \otimes \epsilon':W \otimes W'
\rightarrow K \otimes K \cong K.$

One interesting thing is that the product operation in a graded $K$-algebra $V$ induces an 
additional structure on $V^*$ and vice versa, making the algebra and the coalgebra symmetric 
with respect to the dualization $i.e.$, the dual of an algebra is a coalgebra and the dual of 
a coalgebra is an algebra (with the dual space: $V^*=Hom(V,R)$)
~\cite{Milnor}, ~\cite{Mac}, ~\cite{Swee}, ~\cite{Kass}. 

\null

Given an algebra V and a coalgebra U, a bilinear map called the {\it convolution}
\footnote{Some authors prefer $\ast$ or $\star$, but we use them in
the definition of the star operation.} $\bullet$ is defined in the set $Hom(U,V)$ of
linear maps by ~\cite{Swee}: \\ 
Let $f$ and $g$ be linear maps belonging to $Hom(U,V)$ then $f \bullet g$
satisfies
\begin{equation}
U \harr{\Psi}{} ~~ U \otimes U ~~ \harr{f \otimes g}{} ~~ V \otimes V ~~ \harr{\varphi}{} ~~ V
\end{equation}
obtaining with this operation the structure of monoid in $Hom(U,V)$. 

\null
   
A graded $K$-module $A = \{ A_n \}$ which is at the same time an algebra
and a coalgebra with the morphisms of $K$-graded modules:
\begin{eqnarray*}
\varphi : A \otimes A \rightarrow A,   &\qquad &    \eta : K \rightarrow A, \\
\Psi : A \rightarrow A \otimes A,  & &    \epsilon: A \rightarrow K,
\end{eqnarray*}
such that

$i) (A, \varphi, \eta)$ it's an algebra with $\eta$ a homomorphism of graded coalgebras.

$ii) (A, \Psi, \epsilon)$ it's a coalgebra with $\epsilon$ a homomorphism of graded algebras.

$iii)$ satisfy the axiom connection, that is the diagram
$$\begin{array}{ c c c c c }
   A\otimes A&\harr{\varphi}{}&A&\harr{\Psi}{}&A\otimes A\\
   \varrow{\Psi \otimes \Psi}{}&& &&\varrowu{\varphi \otimes \varphi}{}\\ 
  A\otimes A\otimes A\otimes A& &\harr{id_A\otimes \tau \otimes id_A}{}&
&A\otimes A\otimes A\otimes A\\
  \end{array}$$ 
is commutative, with $id_A\otimes \tau \otimes id_A(a \otimes b \otimes c
\otimes d) = (-1)^{deg~b \cdot deg~c}(a \otimes c \otimes b \otimes d)$,
is called a graded $K$-bialgebra.

Finally a {\it graded Hopf algebra} $A$, is a graded $K$-bialgebra with
a bijective map $S: A \rightarrow A$  called the {\it antipode} which satisfies:
\begin{equation}
\label{7.100}
S \bullet id = id \bullet S = \eta \circ \epsilon
\end{equation} 
and which is also an algebra antihomomorphism, {\it i.e.} the diagram
\begin{equation}
\label{7.200}
\begin{array}{c c c}
A \otimes A & \harr{\varphi}{} & A \\
\varrow{S \otimes S}{} \\
A \otimes A & & \varrow{S}{} \\
\varrow{\tau}{} \\
A \otimes A & \harr{\varphi}{} & A \\
\end{array}
\end{equation}
commutes ~\cite{Milnor}, ~\cite{Tak}. 

Is easy to show that if $A$ has an antipode, it is unique ~\cite{Kass}.
Thus a Hopf algebra is a set of six $(A,\varphi,\eta,\Psi,\epsilon,S)$
satisfiying the axioms above.

\noindent And again the dual of a Hopf algebra has the structure of a Hopf
algebra. 

\null

With the definition of the convolution and the existence of the 
antipode we can form not only a monoid but also a group over $Hom(A,A)$,
since this tell us the form of the inverse elements under $\bullet$. 

\null

A {\it differential Hopf algebra} is a pair $(A,D)$, where $A$ is a Hopf algebra and $D:A \rightarrow A$ is a
differential $(D^2 = 0)$ of degree $\pm 1$ such that the maps, product and co-product $\varphi : A \otimes A 
\rightarrow A$ and $\Psi : A \rightarrow A \otimes A$ are maps of differential modules; $i.e.$ the next
diagram 

\begin{equation}
\begin{array}{ c c c c c }   
   A &\harr{\Psi}{} &A \otimes A& \harr{\varphi}{}&A\\
   \varrow{D}{}&&\varrow{D}{}&&\varrow{D}{}\\
   A &\harr{\Psi}{} &A \otimes A& \harr{\varphi}{}& A\\
   \end{array}
\end{equation}   

\noindent commute.
   
\noindent ($A \otimes A$ has the differential product $D(x \otimes y) = D(x) \otimes y + (-1)^{deg(x)} x \otimes
D(y)$). 
Following with the duality, we finish pointing that the dual of the pair $(A,D)$ is also a
differential Hopf algebra $(A^*,D^*)$.

Now, with all this in mind we can take as our commutative $K$-ring the set of all the 
functions infinitly differentiable from the manifold to the real numbers, $C^{\infty}(M_n)$ and
identify $\Lambda$ with the graded $C^{\infty}(M_n)$-module (now, all the
tensorial products are
over $C^{\infty}(M_n)$, then $\otimes$ is equal to ${\otimes}_{C^{\infty}(M_n)}$).
Then we use the external multiplication $\wedge$ as the product according
with: $\varphi(\alpha \otimes \beta) 
= \alpha \wedge \beta$, then since $\wedge$ is associative (\ref{2.9,1}) 
we see that with $\varphi (\alpha \otimes \varphi(\beta \otimes \gamma)) =
\varphi (\varphi (\alpha \otimes \beta)
\otimes \gamma)= \alpha \wedge \beta \wedge \gamma$ and with unit $1$ (the
constant function $1$ in $C^{\infty}(M_n)$) the set $(\Lambda, \varphi,1)$
form an algebra, and even more a commutative one.

To form a coalgebra we define the $coproduct$ over $\Lambda$ as follows
\begin{equation}
\Psi (\beta) = \sum \Psi (\beta_p) ~~~{\rm with}~\beta \in \Lambda ~{\rm and}~ 
\beta_p \in \Lambda^p,
\end{equation}
and since the map is linear, we only need to define it for the generators. Thus
\begin{equation}
\label{7.800}
\Psi (\alpha_p) = \alpha_p \otimes 1 + 1 \otimes \alpha_p + \sum_{i}
{\alpha^{'}}_i \otimes {\alpha^{''}}_i,
\end{equation}
with $\alpha_p$ being a generator in $\Lambda^p$, and the sum in (\ref{7.800}) defined in such a manner
that the diagram {\it iii}) commutes, (compare with
~\cite{Milnor},~\cite{Mac}).
If we were working with polynomials instead of forms the order in the tensorial
product not matter and all the elements could be added and factorized in the way that exactly becomes the
number ${(^{p}_i})$ in the sumatorial.

Note that each element in the above expression preserve the grade (identifying this grade with the
degree of a form) since $C^{\infty}(M_n) \otimes \Lambda^p \cong \Lambda^p \cong \Lambda^p \otimes C^{\infty}(M_n)$.
The $counit~ \epsilon$ acts in similar way on $\Lambda$
\begin{equation}
\epsilon (\alpha) = \sum \epsilon (\alpha^p), 
\end{equation}
defined over the generators like follows
\begin{equation}
\epsilon(\beta)=\beta ~~ {\rm with} ~~ \beta \in \Lambda^0  ~~ {\rm and}
~~ \epsilon(\alpha)=0 ~~ {\rm in~other~case}
\end{equation}

This is obtained due to (\ref{7.7}) for elements in $\Lambda^0$ and that
from {\it iii}) the elements in $\Lambda^1$ are of the form $1 \otimes
\alpha + \alpha \otimes 1$ so $\epsilon(\alpha)=0$ ~\cite{Kass}, then by 
$\epsilon(\alpha \wedge \beta) = \epsilon(\alpha) \wedge \epsilon(\beta)$
for $\alpha$ and $\beta$ in $\lambda$ and both of degree different of 0,
the action of $\epsilon$ is $0$.  
In this form the triplet $(\Lambda, \Psi, \epsilon)$ defines a cocommutative coalgebra.
We define $\varphi,~\Psi$ and $\epsilon$ in $p$-forms acting over the
elements of the base $dx^{i_1} \wedge ... \wedge dx^{i_p}$ but 
remembering $e^a = {e^a}_\mu dx^\mu$ and in special (\ref{1.5}) when
$ds^2$ is defined, is equivalent to take the definition over $e^{i_1} \wedge ... \wedge e^{i_p}$
because the difference is only in elements of $C^{\infty}(M_n)$.
It is easy to form a bialgebra for the low dimension case, but since
$\Delta$ and $\epsilon$ are morphisms of algebras and $\varphi$ and $1$ are morphisms of coalgebras we only need the form in the generators of $\Lambda^1$ and then extend to the other degrees. 

\null

With the above definition of $\varphi$ and $\Psi$ and with the connection axiom, one easily obtains that the set
$(\Lambda, \varphi, 1, \Psi, \epsilon)$ forms a bialgebra; we can endow this set with an antipode map just using the calculus for the generators following the condition (\ref{7.100}) for the elements of $C^{\infty}(M_n)$ and in $\Lambda^1$, all the others

 can be generated in a recursive way 
\begin{equation}
  S(\alpha) = - \alpha - \sum {\alpha^{'}}_i \wedge S({\alpha^{''}}_i),
\end{equation}
since $\eta \circ \epsilon = 0$ for elements of degree greater than $0$, 
with this and from (\ref{7.200}) $S(\alpha \wedge \beta) = (-1)^{deg~\alpha \cdot deg~\beta} S(\beta) \wedge S(\alpha)$ we obtain 
\begin{equation}
S(dx^{i_1} \wedge ... \wedge dx^{i_p}) = (-1)^p dx^{i_1} \wedge ... \wedge
dx^{i_p}.
\end{equation}
Using now the differential operations defined in the past sections: $d$ of degree $+1$ and $\delta$ of degree $-1$; we can replace respectively in the next diagrams

\begin{equation}
\begin{array}{ c c c }   
   \Lambda \otimes \Lambda& \harr{\varphi}{}&\Lambda\\
   \varrow{d}{}& &\varrow{d}{}\\
   \Lambda \otimes \Lambda& \harr{\varphi}{}& \Lambda\\
   \end{array}
\end{equation} 
  
\begin{equation}
\begin{array}{ c c c }   
   \Lambda \otimes \Lambda& \harr{\varphi}{}&\Lambda\\
   \varrow{\delta}{}&&\varrow{\delta}{}\\
   \Lambda \otimes \Lambda& \harr{\varphi}{}& \Lambda\\
   \end{array}
\end{equation}   

\noindent such that $d$ and $\delta$ are morphisms of algebras (not necessarily a diferential Hopf algebras).
[In the next part of our work we consider the Hopf algebra connected with the $\buildrel{*}\over{\wedge}$ product].

\begin{center}
\section{The harmonic operator}
\end{center}
\setcounter{equation}{0}

The harmonic operator $\Delta$ is a map
\begin{equation}
\label{6.1}
 \Delta : \Lambda^p \rightarrow \Lambda^p
\end{equation}
defined as a composition of the maps previously considered
\begin{equation}
\label{6.2}
\alpha \in \Lambda^p \rightarrow \Delta\alpha := (d\delta + \delta d)\alpha 
\in \Lambda^p.
\end{equation}
Of course, $\Delta$ can be also interpreted as the map $\Lambda \rightarrow \Lambda$ defined by the ``partial maps"
as $\Delta \alpha = \sum^n_{p=0} \Delta \alpha_p$.
The $\Delta$ operator commutes with the basic operators previously considered:
\begin{eqnarray}
\label{6.3}
deg \alpha = p   \rightarrow *\Delta \alpha & = & \Delta * \alpha = (-1)^p (\delta * \delta - d*d) \alpha \\
                             d\Delta \alpha & = &  \Delta d \alpha = d\delta d\alpha \nonumber \\ 
                             \delta \Delta \alpha & = & \Delta \delta \alpha = \delta d \delta \alpha. \nonumber
\end{eqnarray}
Now, from equation (\ref{5.3})
\begin{eqnarray}
\label{6.4}
\alpha, \beta \in \Lambda^p \rightarrow d(\beta \wedge *d \alpha) & = & -\delta d\alpha \wedge *\beta + d\alpha \wedge *d\beta \\
                                        d(\delta \alpha \wedge * \beta) & = & d\delta \alpha \wedge *\beta - 
                                        \delta \alpha \wedge * \delta \beta. \nonumber
\end{eqnarray}
This implies
\begin{equation}
\label{6.4a}
\alpha, \beta \in \Lambda^p \rightarrow \Delta \alpha \wedge * \beta = d\alpha \wedge *d\beta + \delta \alpha \wedge *
\delta \beta + d(\delta \alpha \wedge * \beta - \beta \wedge *d\alpha).
\end{equation}
The co-image of this $\Lambda^n$ equality is a $\Lambda^0$ identity
\begin{equation}
\label{6.4b}
\alpha, \beta \in \Lambda^p \rightarrow *\beta \buildrel{*}\over{\wedge} \Delta \alpha = *d\beta \buildrel{*}\over{\wedge}
d\alpha + *\delta \beta \buildrel{*}\over{\wedge} \delta \alpha + \delta (-1)^p (*\delta \alpha \buildrel{*}\over{\wedge}
\beta + *\beta \buildrel{*}\over{\wedge} d\alpha). 
\end{equation}
Using the fact that for $\alpha, \beta \in \Lambda^p, \alpha \wedge *\beta = \beta \wedge *\alpha, 
*\alpha \buildrel{*}\over{\wedge} \beta = *\beta \buildrel{*}\over{\wedge} \alpha$, one easily finds that
(\ref{6.4a}), (\ref{6.4b}) imply the generalized Green's formula valid for every $\alpha,\beta \in \Lambda^p$
\begin{equation}
\label{6.5a}
\Delta \alpha \wedge *\beta - \Delta \beta \wedge *\alpha = d(\delta \alpha \wedge *\beta - \beta \wedge * d\alpha -
\delta \beta \wedge *\alpha + \alpha \wedge * d\beta),
\end{equation}
\begin{equation}
\label{6.5b}
*\beta \buildrel{*}\over{\wedge} \Delta \alpha - *\alpha \buildrel{*}\over{\wedge} \Delta \beta = \delta (-1)^p (*\delta 
\alpha \buildrel{*}\over{\wedge} \beta + * \beta \buildrel{*}\over{\wedge} d\alpha - * \delta \beta \buildrel{*}\over{\wedge}
\alpha - *\alpha \buildrel{*}\over{\wedge} d\beta).
\end{equation}
Assume now that: \\ 
$a$) The basic manifold $M_n$ is compact, \\
$b$) The signature of the Riemannian metric is positive definite (i.e., $n_{(-)}=0,
~n_{(+)}=n$).

Then the space of $p$-forms $\Lambda^p$ has the structure of the real Hilbert space with a definite positive scalar
product ~\cite{Lov},~\cite{Guille}.
Indeed, one can then define in $\Lambda^p$ the symmetric scalar product
\begin{eqnarray}
\label{6.6}
\alpha , \beta \in \Lambda^p: (\alpha,\beta) = (\beta,\alpha) & := & \int_{M_n} \alpha \wedge * \beta = \int_{M_n} *(*\alpha 
\buildrel{*}\over{\wedge} \beta), \\
 & = & p! \int_{M_n} \alpha_{a_1 ... a_p} \beta^{a_1 ... a_p} *1. \nonumber
\end{eqnarray}
The compactness of $M_n$ is needed to guarantee the existence of the integral (\ref{6.6}) and to assure
from the Gauss theorem
\begin{equation}
\label{6.7}
\alpha \in \Lambda^{n-1} \rightarrow \int_{M_n} d\alpha = 0.
\end{equation}
The positive definite signature of $V_n$ is needed to guarantee that $\alpha_{a_1 ... a_p}\alpha^{a_1...a_p} \geq 0$
and consequently
\begin{equation}
\label{6.8}
(\alpha,\alpha) = 0 \leftrightarrow \alpha = 0.
\end{equation}
Of course, $\Lambda = \bigoplus^n_{p=0} \Lambda^p$ can be also interpreted as a Hilbert space: if $\alpha = 
\sum^n_{p=0} \alpha_p, \beta = \sum^n_{p=0}\beta_p \in \Lambda$, then one understands as the scalar product in 
$\Lambda: (\alpha,\beta)=\sum^n_{p=0}(\alpha_p,\beta_p)$, with $(\alpha_p,\beta_p)$ defined by (\ref{6.6}). 

Now, from (\ref{5.3}) and (\ref{6.7}) one easily finds that for every $\alpha,\beta \in \Lambda$
\begin{equation}
\label{6.9}
(d\alpha,\beta) = (\alpha,\delta \beta).
\end{equation}
Therefore, $\delta$ and $d$ represent a pair of linear operators on $\Lambda$ which are conjugated in the 
sense of ($\cdot,\cdot$). Of course, (\ref{6.9}) can be also interpreted as
\begin{equation}
\label{6.10}
(d\alpha_{p-1},\beta_p) = (\alpha_{p-1},\delta\beta_p),
\end{equation}
an equality for the forms of the definite degrees; the left hand member is here the scalar product in the sense of
$\Lambda^p$, the right hand member is the scalar product in the sense of $\Lambda^{p-1}.$
Notice that (\ref{6.9}) and the nilpotence of $d$ and $\delta$ imply that for every $\alpha,\beta \in \Lambda$:
\begin{equation}
\label{6.11}
(d\alpha,\delta \beta) = 0.
\end{equation}  
This orthogonality of differentials and co-differentials in $\Lambda$ represents a very useful property; some of its
consequences will be discussed later.
Quite similarly as in the case of (\ref{6.9}), applying (\ref{6.4a}) and (\ref{6.7}) one easily finds that for every $\alpha, \beta \in
\Lambda$
\begin{equation}
\label{6.12a}
(\Delta \alpha, \beta) = (d\alpha, d\beta) + (\delta \alpha, \delta \beta) = (\alpha, \Delta \beta).
\end{equation}
Therefore, $\Delta$ is a self-conjugated operator on $\Lambda$; because (\ref{6.12a}) results from the more specific
\begin{equation}
\label{6.12b}
(\Delta \alpha_p, \beta_p) = (d\alpha_p, d\beta_p) + (\delta \alpha_p,\delta \beta_p) = (\alpha_p, \Delta \beta_p),
\end{equation}
$\Delta$ is also a self-conjugated operator on $\Lambda^p$. Notice also that from (\ref{6.12a}) or
(\ref{6.12b}) and $\alpha \neq 0 \rightarrow (\alpha,\alpha) > 0$
\begin{equation}
\label{6.13}
(\Delta \alpha,\alpha) \geq 0,
\end{equation}
so that the operator $\Delta$ is elliptic [in both senses, as the operator on whole $\Lambda$ or on the specific
$\Lambda^p$; parallel to (\ref{6.13}) we have $(\Delta \alpha_p,\alpha_p) \geq 0$].

A form $\alpha$ such that $\Delta \alpha = 0$ is called harmonic (this is the definition of a harmonic form when
$a$ and $b$ are assumed). It easily follows from (\ref{6.12a}) or (\ref{6.12b}) that
\begin{equation}
\label{6.14}
\Delta \alpha = 0 ~~ \leftrightarrow ~~ d\alpha = 0, ~~ \delta \alpha = 0.
\end{equation}
The harmonic form can posses the degree determined (say, $deg \alpha = p$) or undetermined; a general harmonic $\alpha
\in \Lambda$ represents a sequence of harmonic forms of determinated degrees.

Consider now the generalized Poisson's equation:
\begin{equation}
\label{6.15}
 \Delta \alpha = \beta,
\end{equation} 
with $\beta$ given (of undetermined or determined degree). Let $\varphi$ be an arbitrary harmonic form; then 
$(\beta,\varphi) = (\Delta \alpha, \varphi) = (\alpha, \Delta \varphi) = 0$.
Thus, (\ref{6.15}) requires as a necessary consistence condition the orthogonality of the source $\beta$ to all harmonic forms.

We will mention an important result of the theory of the harmonic forms called the $Hodge ~ theorem$ ~\cite{Nak},~\cite{Schwarz}. The theorem states that an arbitrary form $\omega \in \Lambda^p$ then: 1).- can be always represented by 
\begin{equation}
\label{6.16}
\omega = d\alpha + \delta \beta + \gamma , \,\,\, \Delta \gamma = 0,
\end{equation}
where $\alpha \in \Lambda^{p-1}, \beta \in \Lambda^{p+1}, \gamma \in \Lambda^p$ and 2).- $\omega$ determines the
forms $d\alpha, \delta \beta$ and the harmonic form $\gamma$ uniquely.
The proof of the existence (global) of such $\alpha, \beta, \gamma$ to a given $\omega$ that
(\ref{6.16}) is valid, is relatively involved and will be not discussed here. The uniqueness of the decomposition
(\ref{6.16}), however, can be simply demonstrated. Indeed, because $\Delta \gamma = 0 \rightarrow d\gamma = 0 = \delta \gamma$, therefore
assuming (\ref{6.16}) and applying (\ref{6.9}), (\ref{6.11}) one easily finds
\begin{equation}
\label{6.17}
(\omega , d\alpha) = (d\alpha, d\alpha), \,\, (\omega, \delta \beta)=(\delta \beta,\delta \beta),  \,\, (\omega, \gamma) =
(\gamma,\gamma).
\end{equation}
Therefore, from (\ref{6.8}): $\omega = 0 \rightarrow d\alpha = 0, \delta \beta = 0, \gamma=0$.
Of course, the forms $\alpha$ and $\beta$ in (\ref{6.16}) are not uniquely determined; we do not change anything by
adding to $\alpha$ a closed form and to $\beta$ a co-closed form respectively.
Notice that when $\omega$ is closed, $d\omega = 0$, then the Hodge theorem gives: $\omega = d\alpha + \gamma,
\Delta \gamma = 0$; indeed, using (\ref{6.16}) and (\ref{6.17}) we have: $(\delta \beta, \delta \beta) = (\omega, \delta
\beta) = (d\omega, \beta) = 0 \rightarrow \delta \beta = 0$.
Similarly, when $\omega$ is co-closed, $\delta \omega = 0$, then $\omega = \delta \beta + \gamma, \Delta
\gamma = 0$.

The power of the Hodge theorem consists in its global nature. Because we are mostly interested in the
applications of the apparatus of forms in general relativity where $n_{(+)} = 1,~n_{(-)} = 3$ and the basic
$M_4$ is not compact, we will not develop here any further the harmonic analysis of forms which is founded on
the assumptions $a, b$. Further on, however, we will consider to what extend some of the main results of
this analysis remain true when $a$ and $b$ are not assumed.
Now, abandoning $a, b$ consider the maps $\Lambda^p \rightarrow \Lambda^p$ defined by the compositions
$d\delta$ and $\delta d$.
When $\alpha \in \Lambda^p$ has the local representation (\ref{2.4}) then the local representation of the discussed
maps amounts to
\begin{equation}
\label{6.18a}
d\delta \alpha = -p {{\alpha_{\mu_1 ... \mu_{p-1} \lambda}}^{;\lambda}}_{;\mu_p} dx^{\mu_1} \wedge ... \wedge dx^{\mu_p}
\end{equation}
and
\begin{equation}
\label{6.18b}
\delta d\alpha = p{{\alpha_{\mu_1 ... \mu_{p-1} \lambda ;\mu_p}}}^{;\lambda} dx^{\mu_1} \wedge ... \wedge dx^{\mu_p}
- {\alpha_{\mu_1 ... \mu_p ;\lambda}}^{;\lambda} dx^{\mu_1} \wedge ... \wedge dx^{\mu_p}. 
\end{equation}
This permits - by the application of the tensorial Ricci formula for the commutator of $``;_\alpha"$
derivatives - to obtain the local representation of $\Delta \alpha, \alpha \in \Lambda^p$
\begin{equation}
\label{6.19}
-\Delta \alpha = {\alpha _{\mu_1 ... \mu_p ;\lambda}}^{;\lambda} dx^{\mu_1} \wedge ... \wedge dx^{\mu_p} +
R[\alpha],
\end{equation}
where $R[\alpha]$ denotes the form
\begin{eqnarray}
\label{6.20}
R[\alpha] & = & (-p \alpha_{\mu_1 ... \mu_{p-1} \rho}R^\rho _{\mu_p} + \frac{1}{2}p(p-1) \alpha_{\mu_1 ... \mu_{p-2} \rho
\sigma} {R^{\rho \sigma}}_{\mu_{p-1} \mu_p}) dx^{\mu_1} \wedge ... \wedge dx^{\mu_p} \nonumber\\
\nonumber  & = & \frac{1}{p!} \alpha_{\nu_1 ... \nu_p}(\frac{1}{2} \delta^{\nu_1 ... \nu_p \rho \sigma}_{\lambda_1 ...
\lambda_p \mu \nu} {R^{\mu \nu}}_{\rho \sigma} - \delta^{\nu_1 ... \nu_p \rho}_{\lambda_1 ...
\lambda_p \mu} {R^{\mu}}_{\rho}) dx^{\lambda_1} \wedge ... \wedge dx^{\lambda_p} \\
\nonumber  & = & \frac{1}{p!} \alpha_{\nu_1 ... \nu_p}\left(\frac{1}{2} \delta^{\nu_1 ... \nu_p \rho \sigma}_{\lambda_1 ...
\lambda_p \kappa \lambda} {C^{\kappa \lambda}}_{\rho \sigma} + \frac{n-2p}{n-2}\delta^{\nu_1 ... \nu_p \rho}_{\lambda_1 ...
\lambda_p \sigma} \overline{R^\sigma _\rho} \right. \nonumber \\
& & \left. \hspace{7em}
 - \frac{p(n-p)}{n(n-1)} \delta^{\nu_1 ... \nu_p}_{\lambda_1 ...\lambda_p}R \right)
dx^{\lambda_1} \wedge ... \wedge dx^{\lambda_P}.
\end{eqnarray}
The last line of (\ref{6.20}) exhibits explicity the mechanism according to which the irreducible parts of the
curvature enter in the expression for $\Delta \alpha$. In the extreme cases of $p = 0,1, n-1, n$
(\ref{6.19}) and (\ref{6.20}) reduce to
\begin{equation}
\label{6.21}
\Lambda^0 \ni \alpha: -\Delta \alpha = {\alpha_{; \lambda}}^{;\lambda},
\end{equation}
\begin{equation}
\Lambda^1 \ni \alpha: -\Delta \alpha = {\alpha_{\mu ; \lambda}}^{;\lambda}dx^{\mu} - \alpha_{\lambda}
R^\lambda_\mu dx^\mu, \nonumber
\end{equation}
\begin{equation}\begin{array}{l}
\Lambda^{n-1} \ni \alpha: -\Delta \alpha = \nonumber \\  \hspace{3ex}
\left({\alpha_{\mu_1...\mu_{n-1};\lambda}}^{;\lambda}
-\frac{1}{(n-1)!} \alpha_{\nu_1 ... \nu_{n-1}} \delta^{\nu_1 ... \nu_{n-1} \rho}_{\mu_1 ... \mu_{n-1} \sigma}
R^\sigma _\rho\right)dx^{\mu_1} \wedge ... \wedge dx^{\mu_{n-1}} 
\end{array}\end{equation}
\begin{equation}
\Lambda^n \ni \alpha: -\Delta \alpha = {\alpha_{\mu_1 ... \mu_n ; \lambda}}^{;\lambda} dx^{\mu_1} \wedge
... \wedge dx^{\mu_n}. \nonumber
\end{equation}
Thus, from the point of view of the tensorial components, for $p$ = 0 and $p$ = $n$, $\Delta$ reduces simply to the
operator $-g^{\mu \nu}\Delta_{\mu}\Delta_{\nu}$, while for $p$ = 1 and $p$ = $n$-1 beside of this operator appear
terms linear in the Ricci tensor and the form considered.
As far as the conformal curvature is concerned, it can appear in $\Delta \alpha$ only when $1).- n \geq 4 \,\,\,
2).- p \neq 0,1,n-1,n.$ In particular, for $n$ = 4 (\ref{6.21}) covers already the cases of $p=0,1,3,4$; in the remaining
case of $p$ = 2 we have
\begin{equation}
\label{6.22}
n = 4, \alpha \in \Lambda^2 \rightarrow - \Delta \alpha = ({\alpha_{\mu_1 \mu_2;\lambda}}^{;\lambda} -
\frac{1}{3}R \alpha_{\mu_1 \mu_2} + \alpha_{\nu_1 \nu_2} {C^{\nu_1 \nu_2}}_{\mu_1 \mu_2})dx^{\mu_1} \wedge
dx^{\mu_2}.
\end{equation}
Notice the absence of $\overline{R^\alpha_\beta}$ in this expression. 
The harmonic operator acting on $e^a$ results in the formula
\begin{equation}
\label{6.23}
 - \Delta e^a = ({{e^a}_{\mu ; \lambda}}^{; \lambda} {e_b}^{\mu} - R^a_b)e^b = (\delta {\Gamma^a}_b - R^a_b -
 \Gamma^{ars} \Gamma_{brs})e^b.
\end{equation}
Thus, the skew part of the $n \times n$ matrix which acts on $e^b$ in this expression coincides with 
$\delta{\Gamma^a}_b$; the object which remains arbitrary in the considered formalism.

\begin{center}
\section{Conclusions}
\end{center}
Using the formalism of differential forms many of the calculations are easier specially for the
requirements of General Relativity, we reduce the number of indices and components and obtain more
structure in the developed space $\Lambda$ thanks to the operations defined on them.

Working with forms reveals the existence of potentials (Poincar\'e's lemma), the
independence from coordinate systems, and the role of the metric.

The Clifford calculus ~\cite{Lawson},~\cite{Chisholm}, can also be adapted to geometric problems specially to
electromagnetism, although its application is not so clear as in the case of the Cartan calculus. The point of
departure is the role of the metric.

The calculus in differential geometry and specially in general relativity is in favor of 
Cartan's calculus aproach than with tensor calculus, but this depends mainly on the
background of the person doing the computation.
Perhaps even in some fields the Clifford aproach is superior, or perhaps a combination of
the two formalisms (Cartan and Clifford) into a single algebra (Kahler-Atiyah ~\cite{Chisholm}) with two
products could be superior to either. In the case of general relativity the Cartan
calculus looks like perfectly adjusted.
In the second part we deal with the complex case which now plays a distinguished role in general relativity (complex relativity, self-dual gravity,..., etc.).

\begin{center}
\section{Acknowledgments}
\end{center}
The authors wish to express his gratitude to Maciej Przanowski for the time inverted in the review and the corrections 
of this work, and for his valuable suggestions and advices.
Also to Jesus Gonzalez and Hector Garc\'{\i}a Compe\'an for the consultants, comments and the many useful discussions.

\renewcommand{\refname}{

\end{document}